# Ollivier-Ricci curvature in cycle overlap mode

Zexian Zhou [1,2†], Bo Jiao [1*†]
23020241154479@stu.xmu.edu.cn jiaoboleetc@outlook.com

[1] *Key Laboratory of Intelligent Manufacturing and Industrial Internet Technology, Fujian Province University, Xiamen University Tan Kah Kee College, Zhangzhou, 363123, Fujian, China.*
[2] *School of Informatics, Xiamen University, Xiamen, 361102, Fujian, China.*

**Abstract:** Ollivier-Ricci curvature of an edge $(x, y)$ is defined by comparing the distance taken to transport from neighbors of $x$ to neighbors of $y$. It is a structural measure that has been studied in many fields such as community detection and deep neural networks. However, high computational complexity or error limits its application on large scale-free graphs. This paper proposes an optimal transport principle to minimize the distance by 3,4,5-cycles that include the edge $(x, y)$, and designs a curvature calculation approach named Curvature in Cycle Overlap Mode (CCOM). In this approach, a greedy and pruning algorithm is proposed to approximate the optimal transport principle. We theoretically and experimentally verified that our approach CCOM can significantly improve the accuracy of the curvature on real-world networks with low time consumption. In addition, we compared CCOM with baseline approximation approaches in community detection tasks using the same curvature-based framework, and experimentally confirmed the effectiveness of CCOM on large scale-free graphs.

## 1 Introduction

Ollivier-Ricci curvature (ORC) [1-5] measures how the neighborhoods of a pair of nodes are structurally related. It is a much more descriptive measure compared to ones that only focus on node specific attributes or limited topological information such as degree [2]. Positive curvature edges are located within communities, while negative curvature edges form bridges connecting different communities; thus, ORC has been widely studied for community detection [6-10]. Very positively curved edges cause over-smoothing, whereas low curvature value characterizes graph bottlenecks that are strongly related to over-squashing [11]; thus, ORC is important for alleviating the issues of over-smoothing and over-squashing in graph neural networks [11-13]. Subgraph sampling that maximizes the edge curvature amounts to minimizing the distributional difference between a training graph and multiple subgraphs [14]; thus, ORC provides a new research direction for minibatch learning based on subgraph sampling [14,15]. In addition, ORC has been widely studied in graph anomaly detection [16], core subgraph detection [17], and many other applications.

One bottleneck limiting ORC applications is its high computational complexity and memory consumption. The exact calculation of ORC on each edge relies on the solution of a linear programming model [1-9,11-13,16], making it difficult to use for graphs with more than 20,000 nodes. The approximate calculation of ORC has been studied. Sinkhorn distances [18] solve the optimal transport problem through Sinkhorn-Knopp iteration, which exhibits good parallelization, but its results depend on the selection of regularization parameters, and its improvement is not significant. The curvature calculation for the edge $(x, y)$ can be transformed into a minimum weight perfect matching (MWPM) problem on a complete bipartite graph [19], which constructs the bipartite graph us-ing the neighbors of $x$ and $y$, and adopts a matching algorithm to solve the problem. The error of MWPM is 0 if $d_x = d_y$, otherwise, its complexity or error increases. Based on optimal transport theory, a particular transport plan that saves transport costs by 3-cycles can be designed for constructing a lower bound of the exact ORC [20]. The lower bound, named localized curvature with 3-cycles (LoCur), is often used as an approximation for the exact ORC [14,20,21], because its error bound can be controlled on Erdös-Rényi random graphs [14,22] and it is computationally efficient. However, the random graphs have degrees with the same expected value, resulting in a uniform distribution of cycles in the graphs. Scale-free graphs with core-periphery structures are abundant [23,24]. Specifically, the large and sparse periphery is densely connected to the small and dense core; thus, most cycles pass through a few centrum nodes of the core [15], resulting in an uneven distribution of cycles. Therefore, the error of the lower bound should be further studied on scale-free graphs. On the basis of the lower bound, calculating an upper bound of the exact ORC and taking their average as an approximation can alleviate the error. However, the average of the lower and upper bounds (ALU) [25] only considers 3-cycle structures. In addition, other curvatures that are not directly related to ORC were proposed, such as Lower Ricci curvature (LRC) [10], Forman Ricci curvature (FRC) [26] and Balanced Forman curvature (BFC) [13,27]. Both LRC and FRC are 3-cycle curvatures, while BFC takes into account both 3-cycle and 4-cycle structures.

[*] Corresponding author.
[†] These authors contributed equally to this work.

ORC is based on the optimal transport theory, thus it is superior in terms of geometric fidelity and semantic richness, although its exact calculation is time-consuming. In large scale-free graphs, the difference of node degrees between the core centrum and the periphery is significant, resulting in a large number of 3,4,5-cycles overlapping each other when passing through the centrum that consists of a few core nodes with top-highest degrees [15]. The contributions of this paper are as follows: (1) we prove the optimal transport principle of ORC based on 3,4,5-cycles, that is, rigorously adhering to the principle leads to the exact ORC; (2) we design an approach CCOM based on the strategies of cycle enumeration, pruning search, and greedy sorting to approximate the principle, and discuss the time complexity and error bound of the approach; (3) we experimentally confirmed that CCOM significantly improves the accuracy of ORC on scale-free graphs with acceptable time complexity, and performs superior in a curvature-based community detection task.

Open-source code is available at *https://github.com/hitreasonx/CCOM.*

The structure of this paper is organized as follows: Section 2 introduces related work, Section 3 proves the optimal transport principle of ORC based on 3,4,5-cycles, Section 4 designs the strategies of cycle enumeration, pruning search, and greedy sorting, and proposes CCOM to approximate the optimal transport principle, Section 5 analyzes the time complexity and error bound of CCOM, and Section 6 demonstrates the superior performance of CCOM in terms of accuracy with acceptable complexity for calculating ORC and executing a community detection task.

# 2 Related work

## 2.1 Definition of ORC

Suppose we have a simple, undirected, and unweighted graph $G = (\mathcal{V}, \mathcal{E})$ with nodes $v \in \mathcal{V}$ and edges $(u, v) \in \mathcal{E}$. Denote the set of neighboring nodes and the degree of node $x \in \mathcal{V}$ as $\mathcal{N}(x)$ and $d_x$, respectively. Then a probability measure $m_x$ at $x$ can be defined as [2]:

$$m_x(p) = \begin{cases} \alpha & \text{if } p = x \\ (1-\alpha)/d_x & \text{if } p \in \mathcal{N}(x) \\ 0 & \text{otherwise} \end{cases} \tag{1}$$

where $\alpha$ is a parameter within $[0,1]$. It is to keep probability mass of $\alpha$ at node $x$ itself and distribute the rest uniformly over the neighborhood. We set $\alpha = 0$ similar to [11,14,15,20,21,25]. Thus, the ORC $\kappa_{xy}$ is defined by 1-Wasserstein distance $W_1(m_x, m_y)$ to geodesic distance $d(x,y)$ [2], where the 1-Wasserstein distance finds the optimal transport plan between two probability measures, and the geodesic distance represents the shortest path length between $x$ and $y$ in graph $G$. Formally, the ORC $\kappa_{xy}$ on edge $(x,y)$ with $d(x,y) = 1$ is defined as [1-15]:

$$\kappa_{xy} = 1 - \frac{W_1(m_x, m_y)}{d(x,y)} = 1 - W_1(m_x, m_y) \tag{2}$$

where $W_1(m_x, m_y)$ is the optimal value of the objective function in the linear optimization problem

$$\begin{aligned} &\text{minimize} && \sum_{p \in \mathcal{N}(x)} \sum_{q \in \mathcal{N}(y)} d(p,q)\pi(p,q) \\ &\text{subject to} && \sum\nolimits_{q \in \mathcal{N}(y)} \pi(p,q) = m_x(p), \text{for } \forall p \in \mathcal{N}(x) \\ & && \sum\nolimits_{p \in \mathcal{N}(x)} \pi(p,q) = m_y(q), \text{for } \forall q \in \mathcal{N}(y) \end{aligned} \tag{3}$$

Note that, $m_y(q) = 1/d_y$ and $m_x(p) = 1/d_x$, for $\forall p \in \mathcal{N}(x), \forall q \in \mathcal{N}(y)$ when $\alpha = 0$, and $\pi(p,q)$ denotes a transport plan that represents the probability mass moving from $p$ to $q$. Given a feasible transport plan $\pi$, we can obtain an upper bound for the exact $W_1(m_x, m_y)$ and therefore a lower bound for the exact $\kappa_{xy}$.

## 2.2 Lower and upper bounds of ORC

The Kantorovich duality of Eq. (3) can be formulated as:

$$\begin{aligned} W_1(m_x, m_y) &= \sup_{f, 1\text{-Lip}} \left[ \sum_{p \in \mathcal{N}(x)} f(p) m_x(p) - \sum_{q \in \mathcal{N}(y)} f(q) m_y(q) \right] \\ &= \sup_{f, 1\text{-}Lip} \left( \frac{1}{d_x} \sum_{p \in \mathcal{N}(x)} f(p) - \frac{1}{d_y} \sum_{q \in \mathcal{N}(y)} f(q) \right) \end{aligned} \tag{4}$$

where $|f(u) - f(v)| \leq d(u,v)$ for $\forall u, v \in V, u \neq v$. Based on the Kantorovich duality, a feasible 1-Lipschitz function $f$ will yield a lower bound for the exact $W_1(m_x, m_y)$ and therefore an upper bound for the exact $\kappa_{xy}$. Some bounds are summarized as follows [14,20,25]:

$$\kappa_{xy} \geq -\left(1 - \frac{1}{d_x} - \frac{1}{d_y} - \frac{\triangle}{\min[d_x, d_y]}\right)_+ - \left(1 - \frac{1}{d_x} - \frac{1}{d_y} - \frac{\triangle}{\max[d_x, d_y]}\right)_+ + \frac{\triangle}{\max[d_x, d_y]} \tag{5}$$

$$\kappa_{xy} \leq \frac{\triangle}{\max[d_x, d_y]} \tag{6}$$

where $\triangle$ denotes the number of 3-cycles including edge $(x,y)$, and $(\cdot)_+$ is defined as $\max[\cdot, 0]$.

3,4,5-cycles can reduce the transport distance $d(p,q)$ of $\pi(p,q)$, whereas cycles with length larger than 5 are ineffective, as shown in Fig. 1. Taking Fig. 1($c$) as an example, we can keep the mass $m_x(6) = \frac{1}{7}$ at node 6 and make it a part of $m_y(6)$ for distance 0, move a mass of $m_x(4) = \frac{1}{7}$ at node 4 to node 8 for distance 1, and move a mass of $m_x(5) = \frac{1}{7}$ at node 5 to node 9 for distance 2. If the optimal transport plan does not go through 4-cycles and 5-cycles, the equality of Eq. (5) holds. LoCur adopts the lower bound of Eq. (5) as an ORC approximation [14]. ALU alleviates errors by taking the average of the lower and upper bounds in Eqs. (5) and (6) [25].

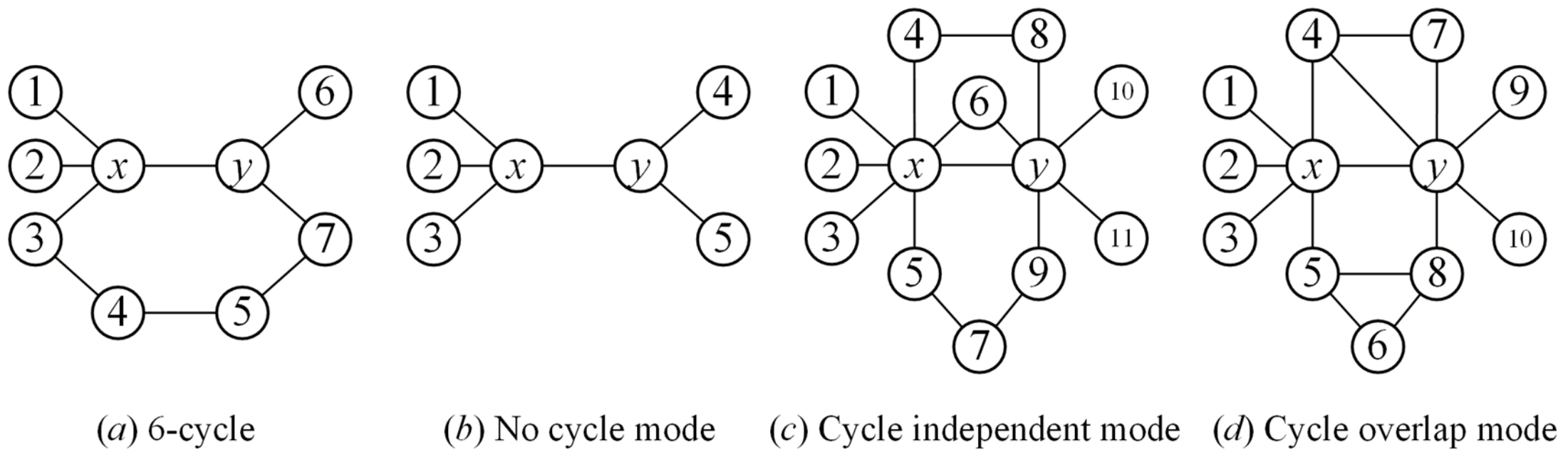

**Figure 1.** Transport plans based on 3,4,5-cycles including edge $(x, y)$. (*a*) 6-cycles cannot reduce the transport distance $d(p, q)$ of $\pi(p, q)$ for $p \in \mathcal{N}(x)$ and $q \in \mathcal{N}(y)$, while 3,4,5-cycles can realize distance reduction. (*b*) $d(p, q) = 3$ in no cycle mode where $p, q \notin \{x, y\}$. (*c*) Except for $x$ and $y$, no node exists in two different 3,4,5-cycles that include edge $(x, y)$. (*d*) Except for $x$ and $y$, there is at least one node that exists in two different 3,4,5-cycles that include edge $(x, y)$.

Curvatures based on 3-cycles are time efficient, which have controllable errors on Erdös-Rényi random graphs [14] with uniform cycle distribution. However, in scale-free graphs, plenty of chains composed of low-degree nodes exist in the large and sparse periphery, and the low-degree nodes in the chains prefer to connect to the small and dense core (especially the core centrum nodes with top highest-degrees). Therefore, 3,4,5-cycles are extensively formed in the scale-free graphs and exhibit an uneven distribution, which leads to the low diameter property [15]. The 3,4,5-cycles tend to pass through the core centrum nodes, resulting in the cycle overlap mode, as shown in Fig. 1(*d*). That is, at least one node $u \in \mathcal{N}(x) \cup \mathcal{N}(y) \wedge u \notin \{x, y\}$ exists in different 3,4,5-cycles including $(x, y)$. In the mode, some 4,5-cycles may be ineffective; for example, the optimal transport plan in Fig. 1(*d*) is as follows: we keep the mass $m_x(4)$ at node 4 as $m_y(4)$ for distance 0, move a mass of $m_x(y)$ at node $y$ to node 7 for distance 1, move a mass of $m_x(1)$ at node 1 to node $x$ for distance 1, move a mass of $m_x(5)$ at node 5 to node 8 for distance 1, move a mass of $m_x(2)$ at node 2 to node 9 for distance 3, and move a mass of $m_x(3)$ at node 3 to node 10 for distance 3. We can observe that the 4-cycle $x - 4 - 7 - y$ and the 5-cycle $x - 5 - 6 - 8 - y$ in Fig. 1(*d*) are ineffective, whereas the 4-cycle $x - 5 - 8 - y$ in Fig. 1(*d*) and the 5-cycle $x - 5 - 7 - 9 - y$ in Fig. 1(*c*) are helpful in reducing the transport cost. Therefore, 4,5-cycles play an important role in approaching the optimal transport plan for scale-free graphs, and we need to prune ineffective 4,5-cycles to alleviate errors and improve time efficiency. BFC [13] considers both 3-cycle and 4-cycle structures, but the design of BFC is not based on the optimal transport theory.

## 2.3 Curvature-based community detection

Many real world networks have community structures, namely groups of well-connected nodes with important functional roles. Curvature-based detection methods include sequential removal of negatively curved edges (SRNCE) [7,28], discrete Ricci flow (DRF) [9,29,30] and curvature-guided graph convolutional network (CGCN) [31]. SRNCE interprets negatively curved edges as “bridges” between communities and cuts them to detect communities. DRF deforms edge curvatures by simulating the evolution process of Ricci flow and naturally emerges community structures. CGCN integrates curvatures with graph convolutional networks for community detection.

This paper proposes an approach CCOM for approximately calculating ORC. Because SRNCE applies curvature values more intuitively, as shown in Fig.2, we choose it to compare the application effects of CCOM with baseline curvature calculation approaches.

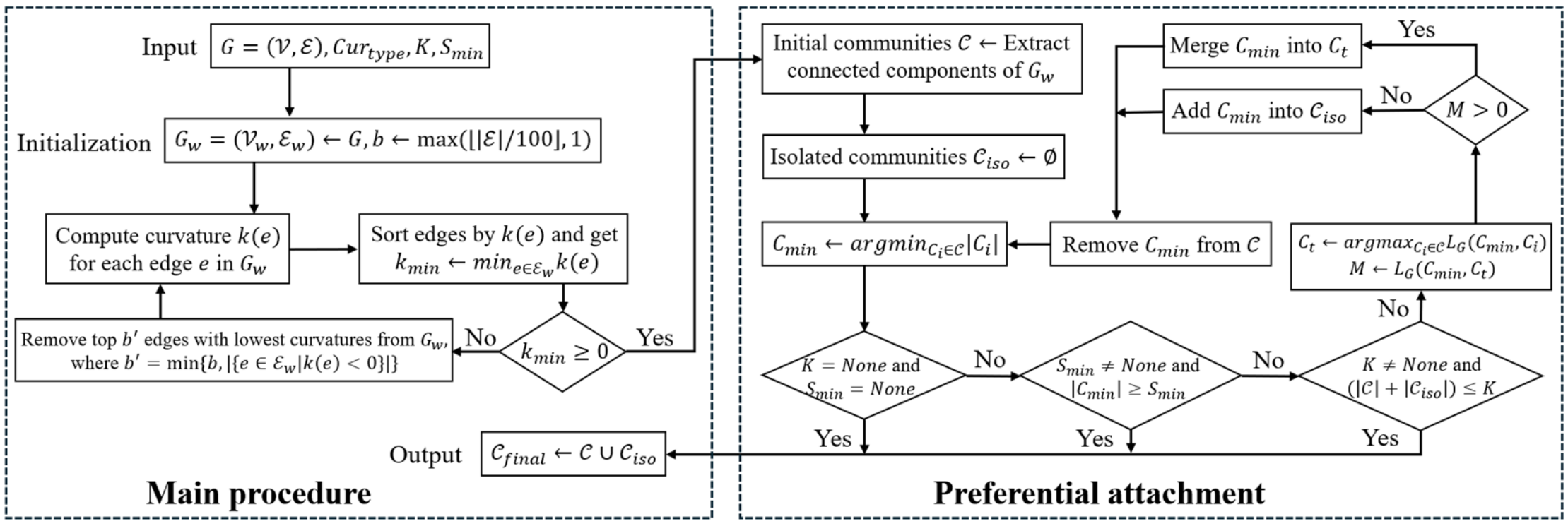

**Figure 2.** Overall flowchart of SRNCE [7], where $Cur_{type}$ represents a chosen type of curvatures, such as CCOM, MWPM, LoCur, ALU, LRC, FRC and BFC, $K$ denotes a predefined number of communities, and $S_{min}$ denotes the size of the smallest community. Note that $L_G(C_{min}, C_i)$ represents the number of edges connecting two communities $C_{min}$ and $C_i$ in original graph $G$, and $|\cdot|$ is defined as the cardinality of a set. To reduce the frequency for repeatedly calculating edge curvatures on large-scale graphs, we add a parameter $b$ to remove more negative-curvature edges at each iteration.

To evaluate the performance of the curvature calculation approaches in community detection tasks, criteria such as Adjusted Rand Index (ARI) [32] and Adjusted Mutual Information (AMI) [33] are commonly used. The criteria consider a community structure as a partition of the node set. Comparing the community structure estimated by SRNCE with the reference community structure therefore consists in comparing two partitions (estimated and reference). ARI evaluates the proportion of node pairs for which both the estimated and reference community structures agree, and ranges from $-1$ (less than chance agreement) to $1$ (complete agreement). AMI compares the two partitions by measuring how much information they have in common, and ranges from $0$ (they are independent) to $1$ (complete agreement). Furthermore, because ARI and AMI are less effective in networks with mixed memberships, we utilize F1-score [34] for evaluation that combines the precision and recall of the detected communities and is particularly useful in networks where a node can be part of multiple communities. The F1-score ranges from $0$ to $1$ (best performance).

# 3 Optimal transport principle

The exact ORC can be derived by constructing particular solutions for Eqs. (3) and (4) in the cycle independent mode that is defined in Fig. 1($c$). The lower bound of Eq. (5) is the exact ORC under the assumption of all 4,5-cycles being ineffective (See supplementary materials that are available at *https://doi.org/10.48550/arXiv.2606.03317*). Based on the proof in supplementary materials, we derive the exact ORC in the cycle independent mode as:

$$\begin{aligned}\kappa_{xy} = &-\left(1 - \frac{1}{d_x} - \frac{1}{d_y} - \frac{\triangle}{\min[d_x, d_y]} - \frac{\blacksquare + \blacklozenge}{\max[d_x, d_y]}\right)_+ \\ &-\left(1 - \frac{1}{d_x} - \frac{1}{d_y} - \frac{\triangle + \blacksquare}{\max[d_x, d_y]}\right)_+ + \frac{\triangle}{\max[d_x, d_y]}\end{aligned} \tag{7}$$

where $\triangle$, $\blacksquare$ and $\blacklozenge$ represent the number of 3-cycles, 4-cycles, and 5-cycles including edge $(x, y)$, respectively, and $(\cdot)_+$ is defined as $\max[\cdot, 0]$.

However, Eq. (7) does not hold in the cycle overlap mode that is defined in Fig. 1($d$). Because it is difficult to construct particular solutions for Eqs. (3) and (4) that are applicable to all cases in the complex mode, we first prove the optimal transport principle of ORC.

Let $\pi: \mathcal{N}(x) \times \mathcal{N}(y) \rightarrow [0,1]$ be a feasible transport plan satisfying

$$\sum_{q\in\mathcal{N}(y)} \pi(p,q) = m_x(p), \text{for } \forall p \in \mathcal{N}(x)$$
$$\sum_{p\in\mathcal{N}(x)} \pi(p,q) = m_y(q), \text{for } \forall q \in \mathcal{N}(y) \tag{8}$$

We define a 4-dimensional vector for the transport plan $\pi$ as:

$$M(\pi) = \left(M_0(\pi), M_1(\pi), M_2(\pi), M_3(\pi)\right) \tag{9}$$

where $M_k(\pi), k = 0,1,2,3$ denotes the sum of the probability mass of $\pi$ transported through a path of length $k$. Formally, we define

$$M_k(\pi) = \sum_{p\in\mathcal{N}(x)\wedge q\in\mathcal{N}(y)\wedge d(p,q)=k} \pi(p,q) \tag{10}$$

Based on the fact that $d(p,q) \in \{0,1,2,3\}$ for $\forall p \in \mathcal{N}(x) \wedge \forall q \in \mathcal{N}(y)$, as shown in Fig. 1, and $\sum_{p\in\mathcal{N}(x)} m_x(p) = \sum_{q\in\mathcal{N}(y)} m_y(q) = 1$, we can derive

$$\sum_{k=0}^{3} M_k(\pi) = 1 \tag{11}$$

We define $\|M(\pi)\|$ as the rank of $M(\pi)$ sorted in lexicographical order for $\forall \pi \in \Omega$ where $\Omega$ denotes the set of all feasible transport plans. Specifically, for $\forall \pi_1, \pi_2 \in \Omega$,

$$\|M(\pi_1)\| > \|M(\pi_2)\| \Leftrightarrow$$
$$\exists k \in \{0,1,2\}, M_k(\pi_1) > M_k(\pi_2) \wedge \forall i < k, M_i(\pi_1) = M_i(\pi_2) \tag{12}$$

It is well-known that $\|M(\pi_1)\| = \|M(\pi_2)\|$ if $\forall i \in \{0,1,2\}, M_i(\pi_1) = M_i(\pi_2)$.

Considering the objective function of Eq. (3), we define the transport cost of $\pi$ as

$$C(\pi) = \sum_{k=0}^{3} k \cdot M_k(\pi) \tag{13}$$

The optimal transport plan corresponds to the minimum transport cost.

**Theorem 1**. $\pi^* = \text{argmax}_{\pi\in\Omega}\|M(\pi)\|$ is the optimal transport plan.

**Proof.** We assume that $\pi' \neq \pi^*$ for the optimal transport plan $\pi' \in \Omega$ that satisfies

$$\pi' = \text{argmax}_{\pi_{opt}}\left\|M(\pi_{opt})\right\|, \text{where } \pi_{opt} = \text{argmin}_{\pi\in\Omega} C(\pi) \tag{14}$$

It is well-known that $\|M(\pi^*)\| > \|M(\pi')\|$. Based on Eq. (12), $\exists k \in \{0,1,2\}$,

$$M_k(\pi^*) > M_k(\pi') \wedge \forall i < k, M_i(\pi^*) = M_i(\pi') \tag{15}$$

Owing to Eq. (11), i.e., $\sum_{i=0}^{3} M_i(\pi^*) = \sum_{i=0}^{3} M_i(\pi')$, based on Eq. (15), we can derive

$$M_k(\pi^*) - M_k(\pi') = \sum_{i=k+1}^{3} [M_i(\pi') - M_i(\pi^*)] \tag{16}$$

Based on Eqs. (13), (15) and (16), we can derive

$$C(\pi^*) - C(\pi') = \sum_{i=k+1}^{3} (i-k) \cdot [M_i(\pi^*) - M_i(\pi')] \tag{17}$$

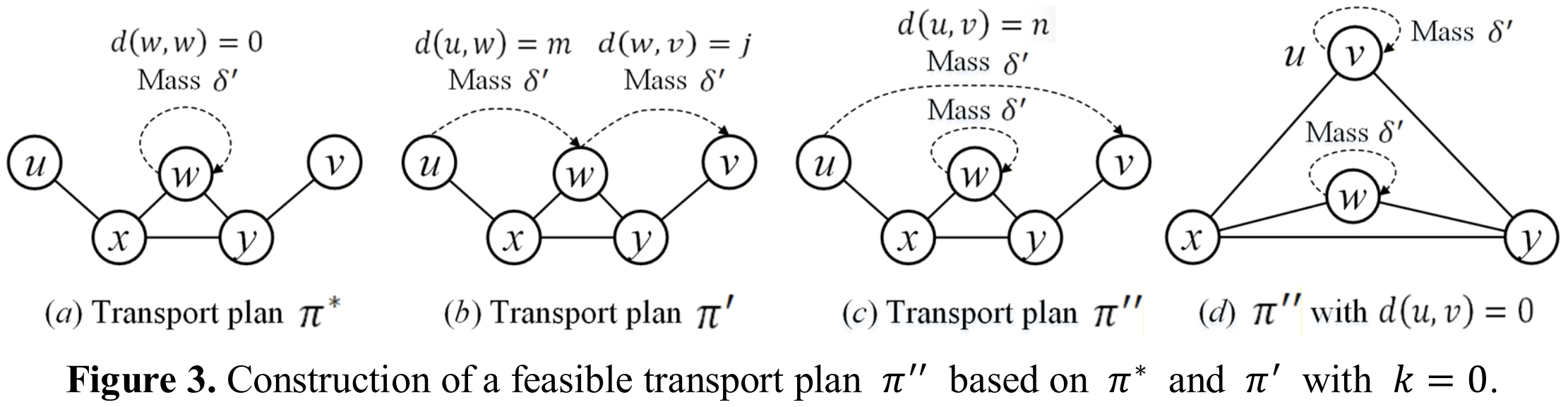

**Figure 3.** Construction of a feasible transport plan $\pi''$ based on $\pi^*$ and $\pi'$ with $k = 0$. The moved mass $\delta'$ is small and positive.

Let $\delta = M_k(\pi^*) - M_k(\pi')$. Then, $\delta > 0$ owing to Eq. (15).

$k = 0$ indicates that $\delta = M_0(\pi^*) - M_0(\pi') > 0$. As shown in Fig. 3, there is a mass of $\delta'$ $(\leq \delta)$ at node $w$ being moved for distance 0 in $\pi^*$, but it does not occur in $\pi'$. Under the constraint of Eq. (8), i.e., $\sum_{q\in\mathcal{N}(y)} \pi(w,q) = m_x(w)$ and $\sum_{p\in\mathcal{N}(x)} \pi(p,w) = m_y(w)$, the mass $\delta'$ at $w$ in $\pi'$ is moved from node $u$ for distance $m$ and to node $v$ for distance $j$, where $m \geq 1 \wedge j \geq 1$.

Then, as shown in Fig.3, we construct a transport plan $\pi''$ as follows:

$$\begin{cases} \pi''(u,v) = \pi'(u,v) + \delta' & d(u,v) = n \\ \pi''(u,w) = \pi'(u,w) - \delta' & d(u,w) = m \\ \pi''(w,v) = \pi'(w,v) - \delta' & d(w,v) = j \\ \pi''(w,w) = \pi'(w,w) + \delta' & d(w,w) = 0 \\ \pi''(p,q) = \pi'(p,q) & \text{otherwise} \end{cases} \tag{18}$$

Using Eq. (8), we can confirm that $\pi''$ is a feasible transport plan.

When $k = 0$, based on Eqs. (10), (13) and (18), we can derive

$$C(\pi'') - C(\pi') = (n - m - j)\delta' \tag{19}$$

Based on the inequality $d(u,w) + d(w,v) \geq d(u,v)$, i.e., $m + j \geq n$, $C(\pi'') - C(\pi') \leq 0$. Because $C(\pi'') < C(\pi')$ contradicts Eq. (14), $C(\pi'') = C(\pi')$, namely, $\pi'' = \mathrm{argmin}_{\pi\in\Omega} C(\pi)$. According to the method in Fig. 3, we observe that

$$M_0(\pi') + \delta' \leq M_0(\pi'') \leq M_0(\pi') + 2\delta' \tag{20}$$

Thus, $M_0(\pi'') > M_0(\pi')$. Based on the definition of Eq. (12), $\|M(\pi'')\| > \|M(\pi')\|$. However, $\pi'' = \mathrm{argmin}_{\pi\in\Omega} C(\pi)$ that contradicts Eq. (14). Therefore, when $k = 0$, we can derive that $\delta = M_0(\pi^*) - M_0(\pi') = 0$, which is contradictory.

Because the same node $w$, where $w \notin \{x, y\}$, cannot exist in different 3-cycles including edge $(x, y)$, 3-cycles are independent of each other. In addition, the method in Fig. 3 can maximize the mass for each 3-cycle. Thus, we confirm that $\pi^*(w,w) = \min[m_x(w), m_y(w)]$ for each 3-cycle $x - w - y$, which is the upper bound of the mass moved for distance 0.

Considering the initial assumption of Eq. (14), Eq. (21) holds.

$$M_0(\pi') = M_0(\pi^*) \text{ and } \|M(\pi')\| \geq \|M(\pi_{opt})\| \text{ for } \forall \pi_{opt} = \mathrm{argmin}_{\pi\in\Omega} C(\pi) \tag{21}$$

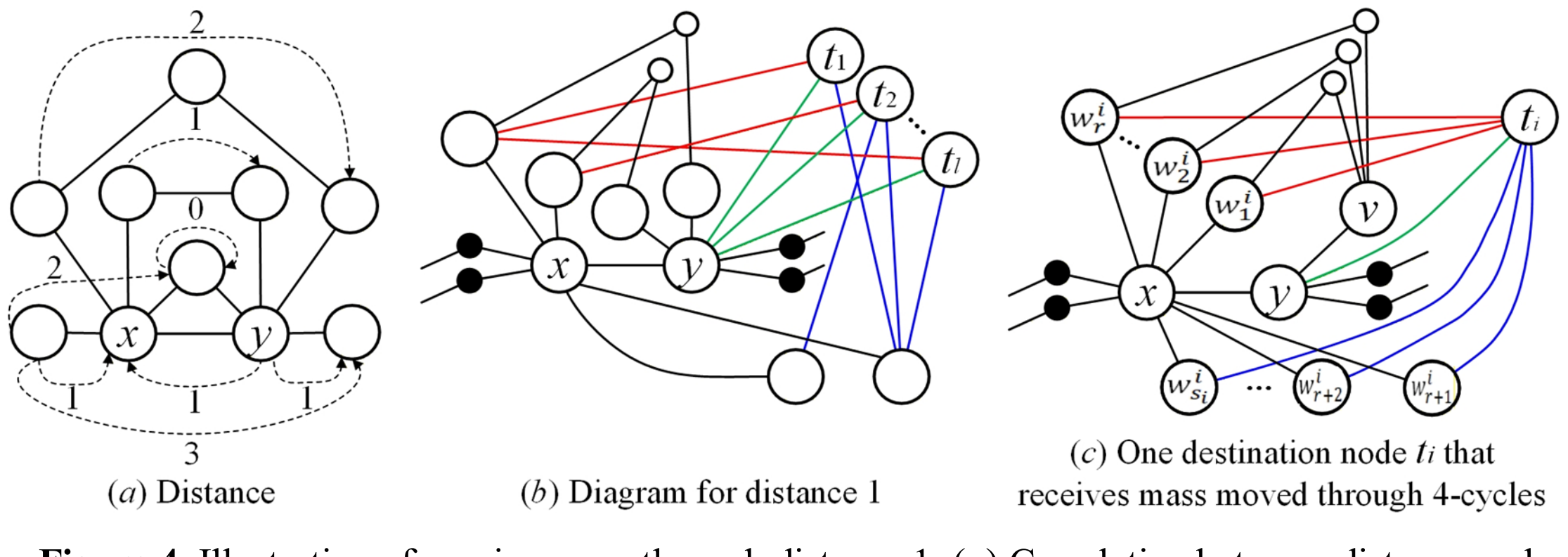

**Figure 4.** Illustration of moving mass through distance 1. ($a$) Correlation between distance and 3,4,5-cycles. ($b$) A diagram that includes all source and destination nodes that transport mass through a distance of 1. ($c$) One destination node $t_i$ $(i = 1,2,\cdots,l)$ in ($b$) that can receive mass through 4-cycles.

Without loss of generality, we assume that

$$d_x \geq d_y \tag{22}$$

for edge $(x, y)$, where $d_x$ and $d_y$ denote the degrees of nodes $x$ and $y$, respectively.

$k = 1$ indicates that $\delta = M_1(\pi^*) - M_1(\pi') > 0$. As shown in Fig. 4($a$), distance 1 corresponds to the transports from neighbors of $x$ to $x$, from $y$ to neighbors of $y$, and from $w$ to $t$ where $x -$ $-w - t - y$ constructs a 4-cycle in graph $G = (\mathcal{V},\ \mathcal{E})$. Fig. 4($b$) exhibits a diagram that includes all the transports through a distance of 1. Define $\{t_i\}_{i=1}^{l} = \{t \in \mathcal{N}(y) \wedge t \neq x | \exists w \in \mathcal{N}(x) \wedge w \neq y, (w, t) \in \mathcal{E}\}$ as the set of all destination nodes that can receive mass through 4-cycles, and define $\{w_1^i, w_2^i, \cdots, w_{s_i}^i\} = \{w \in \mathcal{N}(x) \wedge w \neq y \wedge w \notin \mathcal{N}(y) | (w, t_i) \in \mathcal{E}\}$ as the set of all source nodes that can send mass to $t_i$ though 4-cycles for each destination node $t_i$, as shown in Fig. 4($c$). Note that, based on Eq. (22), the mass $m_x(w)$ was kept at $w$ for distance 0 if $w \in \mathcal{N}(x) \cap \mathcal{N}(y)$.

We first prove two properties of node $t_i$ in Fig. 4($c$), and then discuss the assumption of $k = 1$ with $M_1(\pi^*) - M_1(\pi') > 0$ in three categories.

**Property 1.** Assuming $\pi'(y, x) = 0$ for any optimal transport plan $\pi'$ that satisfies Eq. (14), if there is a node $w_r^i$, $1 \leq r \leq s_i$, satisfying $\exists v \in \mathcal{N}(y), \pi'(w_r^i, v) > 0 \wedge d(w_r^i, v) = j \geq 2$, or if there is a node $w_r^i$, $1 \leq r \leq s_i$, satisfying $\pi'(w_r^i, x) > 0$, then the following equation holds

$$\pi'(y, t_i) = 0 \wedge \sum_{r=1}^{s_i} \pi'(w_r^i, t_i) = m(t_i) \tag{23}$$

where $m(t_i) = m_y(t_i) - \mathbb{1}_{t_i \in \mathcal{N}(x)} \cdot m_x(t_i)$, and $\mathbb{1}_{t_i \in \mathcal{N}(x)} = 1$ if $t_i \in \mathcal{N}(x)$ otherwise 0.

**Proof of Property 1.** If there is a node $w_r^i, 1 \leq r \leq s_i$, satisfying $\exists v \in \mathcal{N}(y), \pi'(w_r^i, v) > 0 \wedge d(w_r^i, v) = j \geq 2$, we assume that $\pi'(y, t_i) > 0$. As shown in Fig. 4($c$), we set $\delta' = \min[\pi'(w_r^i, v), \pi'(y, t_i)]$, and construct a new feasible transport plan $\pi''$ as follows:

$$\begin{cases} \pi''(y, v) = \pi'(y, v) + \delta' & d(y, v) = 1 \\ \pi''(y, t_i) = \pi'(y, t_i) - \delta' & d(y, t_i) = 1 \\ \pi''(w_r^i, v) = \pi'(w_r^i, v) - \delta' & d(w_r^i, v) = j \\ \pi''(w_r^i, t_i) = \pi'(w_r^i, t_i) + \delta' & d(w_r^i, t_i) = 1 \\ \pi''(p, q) = \pi'(p, q) & \text{otherwise} \end{cases} \tag{24}$$

Based on Eqs. (10), (13) and (24), $C(\pi'') - C(\pi') = (1 - j) \cdot \delta' < 0$ that contradicts Eq. (14). Thus, we can confirm that $\pi'(y, t_i) = 0$.

We assume that $\exists u \in \mathcal{N}(x), \pi'(u, t_i) > 0 \wedge d(u, t_i) = m \geq 2$, and set $\delta' = \min[\pi'(w_r^i, v), \pi'(u, t_i)]$. Then, we construct a new feasible transport plan $\pi''$ as follows:

$$\begin{cases} \pi''(u, v) = \pi'(u, v) + \delta' & d(u, v) = n \\ \pi''(u, t_i) = \pi'(u, t_i) - \delta' & d(u, t_i) = m \\ \pi''(w_r^i, v) = \pi'(w_r^i, v) - \delta' & d(w_r^i, v) = j \\ \pi''(w_r^i, t_i) = \pi'(w_r^i, t_i) + \delta' & d(w_r^i, t_i) = 1 \\ \pi''(p, q) = \pi'(p, q) & \text{otherwise} \end{cases} \tag{25}$$

Based on Eqs. (10), (13) and (25), $C(\pi'') - C(\pi') = (1 - j + n - m) \cdot \delta'$. Fig. 4($a$) shows that $n \leq 3$. Because $j \geq 2 \wedge m \geq 2$, when $n \leq 2$ or $j + m > 4$, we derive $C(\pi'') - C(\pi') < 0$, which contradicts Eq. (14). When $n = 3 \wedge j = 2 \wedge m = 2$, we derive that $C(\pi'') = C(\pi')$, namely, $\pi'' = \mathrm{argmin}_{\pi \in \Omega} C(\pi)$. Based on Eqs. (10) and (25), $M_0(\pi'') = M_0(\pi')$ and $M_1(\pi'') = M_1(\pi') + \delta'$, namely $||M(\pi'')|| > ||M(\pi')||$, which contradicts Eq. (21). Therefore, we confirm that $\pi'(u, t_i) > 0$ only if $u \in \mathcal{N}(x) \wedge d(u, t_i) \leq 1$. Because $\pi'(y, t_i) = 0$, $d(u, t_i) = 0$ if and only if $u = t_i$, the mass $m_x(t_i)$ was kept at $t_i$ for distance 0 if $t_i \in \mathcal{N}(x)$ and $d_x \geq d_y$, $d(w_r^i, t_i) = 1$ for $r = 1, 2, \cdots, s_i$, and $m_y(t_i) = \sum_{u \in \mathcal{N}(x)} \pi'(u, t_i)$, we derive that Eq. (23) holds.

If there is a node $w_r^i$, $1 \leq r \leq s_i$, satisfying $\pi'(w_r^i, x) > 0$, we assume that $\pi'(y, t_i) > 0$. As shown in Fig. 4($c$), we set $\delta' = \min[\pi'(w_r^i, x), \pi'(y, t_i)]$ and construct a new feasible transport plan $\pi''$ as follows: $\pi''(y, t_i) = \pi'(y, t_i) - \delta'$, $\pi''(y, x) = \pi'(y, x) + \delta'$, $\pi''(w_r^i, x) = \pi'(w_r^i, x) -$ $\delta'$, $\pi''(w_r^i, t_i) = \pi'(w_r^i, t_i) + \delta'$, and $\pi''(p,q) = \pi'(p,q)$ otherwise. Because $d(y,t_i) = d(y,x) = d(w_r^i, x) = d(w_r^i, t_i) = 1$, we derive that $C(\pi'') = C(\pi')$ and $M_i(\pi'') = M_i(\pi')$ for $i = 0,1,2,3$, that is, $\pi'' = \mathrm{argmin}_{\pi\in\Omega} C(\pi)$ and $\|M(\pi'')\| = \|M(\pi')\|$. Thus, the optimal transport plan $\pi''$ satisfies Eq. (14), namely $\pi'' = \mathrm{argmax}_{\pi_{opt}} \|M(\pi_{opt})\|$ where $\pi_{opt} = \mathrm{argmin}_{\pi\in\Omega} C(\pi)$. Based on the assumption of Property 1, $\pi''(y,x) = 0$. However, $\pi''(y,x) = \pi'(y,x) + \delta' > 0$, which is contradictory. Thus, we can confirm that $\pi'(y,t_i) = 0$.

We assume that $\exists u \in \mathcal{N}(x), \pi'(u,t_i) > 0 \wedge d(u,t_i) = m \geq 2$, and set $\delta' = \min[\pi'(w_r^i, x), \pi'(u,t_i)]$. Then, we construct a new feasible transport plan $\pi''$ as follows: $\pi''(u,x) = \pi'(u,x) + \delta'$, $\pi''(u,t_i) = \pi'(u,t_i) - \delta'$, $\pi''(w_r^i, x) = \pi'(w_r^i, x) - \delta'$, $\pi''(w_r^i, t_i) = \pi'(w_r^i, t_i) + \delta'$, and $\pi''(p,q) = \pi'(p,q)$ otherwise. Because $d(u,x) = d(w_r^i, x) = d(w_r^i, t_i) = 1$, we derive $C(\pi'') - C(\pi') = (1-m)\cdot\delta' < 0$ that contradicts Eq. (14). Therefore, we confirm that $\pi'(u,t_i) > 0$ only if $u \in \mathcal{N}(x) \wedge d(u,t_i) \leq 1$. Because $\pi'(y,t_i) = 0$, $d(u,t_i) = 0$ if and only if $u = t_i$, the mass $m_x(t_i)$ was kept at $t_i$ for distance 0 if $t_i \in \mathcal{N}(x)$ and $d_x \geq d_y$, $d(w_r^i, t_i) = 1$ for $r = 1,2,\cdots,s_i$, and $m_y(t_i) = \sum_{u\in\mathcal{N}(x)} \pi'(u,t_i)$, we derive that Eq. (23) holds.

**Property 2.** If $\sum_{r=1}^{s_i} \pi'(w_r^i, t_i) < m_y(t_i) - \mathbb{1}_{t_i\in\mathcal{N}(x)} \cdot m_x(t_i)$, and $\pi'(y,x) = 0$ for any optimal transport plan $\pi'$ that satisfies Eq. (14), the following equation holds

$$m_x(w_r^i) = \sum_{(w_r^i, t)\in\mathcal{E}\wedge t\in\{t_1,t_2,\cdots,t_l\}} \pi'(w_r^i, t) \quad \text{for } \forall r = 1,2,\cdots,s_i \tag{26}$$

Property 2 can be proved by the contrapositive of Property 1. Based on Properties 1 and 2, we discuss the assumption of $k = 1$ with $M_1(\pi^*) - M_1(\pi') > 0$ in three categories.

**Category 1.** $\pi'(y,x) > 0$ for an optimal transport plan $\pi'$ that satisfies Eq. (14).

For each $v \in \mathcal{N}(y) \wedge v \neq x$, we assume that $\exists u \in \mathcal{N}(x), \pi'(u,v) > 0 \wedge d(u,v) = m \geq 2$, and set $\delta' = \min[\pi'(y,x), \pi'(u,v)]$. Then, we construct a new feasible transport plan $\pi''$ as follows: $\pi''(u,x) = \pi'(u,x) + \delta'$, $\pi''(u,v) = \pi'(u,v) - \delta'$, $\pi''(y,v) = \pi'(y,v) + \delta'$, $\pi''(y,x) = \pi'(y,x) - \delta'$, and $\pi''(p,q) = \pi'(p,q)$ otherwise. Because $d(u,x) = d(y,v) = d(y,x) = 1$, we derive $C(\pi'') - C(\pi') = (1-m)\cdot\delta' < 0$ that contradicts Eq. (14). Thus, $\pi'(u,v) > 0$ only if $u \in \mathcal{N}(x) \wedge d(u,v) \leq 1$, and $m_y(v) = \sum_{u\in\mathcal{N}(x)\wedge d(u,v)\leq 1} \pi'(u,v)$. Based on Eq. (8), $m_y(x) = \sum_{u\in\mathcal{N}(x)} \pi'(u,x)$ where $d(u,x) = 1$ for $\forall u \in \mathcal{N}(x)$. Thus, $M_0(\pi') + M_1(\pi') = \sum_{v\in\mathcal{N}(y)} m_y(v) = 1$. Because $M_0(\pi') = M_0(\pi^*)$, we derive $M_1(\pi^*) \leq M_1(\pi')$ that contradicts the assumption of $M_1(\pi^*) - M_1(\pi') > 0$.

**Category 2.** $\pi'(y,x) = 0$ and Property 1 holds for each destination node $t_i$ with $1 \leq i \leq l$.

Owing to Property 1, $\sum_{i=1}^{l}\sum_{r=1}^{s_i} \pi'(w_r^i, t_i) = \sum_{i=1}^{l} m(t_i)$ that is the upper bound of the mass moved through 4-cycles. Thus, $\sum_{i=1}^{l}\sum_{r=1}^{s_i} \pi'(w_r^i, t_i) \geq \sum_{i=1}^{l}\sum_{r=1}^{s_i} \pi^*(w_r^i, t_i)$.

Based on Eq. (8), $\sum_{u\in\mathcal{N}(x)} \pi'(u,x) = \sum_{u\in\mathcal{N}(x)} \pi^*(u,x) = m_y(x)$ that is the total mass moved from neighbors of $x$ to $x$, and $\sum_{v\in\mathcal{N}(y)} \pi'(y,v) = \sum_{v\in\mathcal{N}(y)} \pi^*(y,v) = m_x(y)$ that is the total mass moved from $y$ to neighbors of $y$. As is well known, $\pi'(y,x)$ is included in both $\sum_{u\in\mathcal{N}(x)} \pi'(u,x)$ and $\sum_{v\in\mathcal{N}(y)} \pi'(y,v)$. Thus, the total mass of $\pi'$ and $\pi^*$, moved for distance 1 and without passing through 4-cycles, is $m_y(x) + m_x(y) - \pi'(y,x)$ and $m_y(x) + m_x(y) - \pi^*(y,x)$, respectively. Owing to the condition $\pi'(y,x) = 0$, $m_y(x) + m_x(y) - \pi'(y,x) \geq m_y(x) + m_x(y) - \pi^*(y,x)$.

Thus, $M_1(\pi') \geq M_1(\pi^*)$, which contradicts the assumption of $M_1(\pi^*) - M_1(\pi') > 0$.

**Category 3.** $\pi'(y,x) = 0$, Property 1 holds for destination nodes $t_i \in T_1$, and Property 2 holds for destination nodes $t_i \in T_2$, where $T_1 \cap T_2 = \emptyset$ and $T_1 \cup T_2 = \{t_1, t_2, \cdots, t_l\}$.

Based on Category 2, $\pi'(y,x) = 0$ indicates that the total mass of $\pi'$ moved for distance 1 and without passing through 4-cycles is $m_y(x) + m_x(y)$, and the total mass is not less than that of $\pi^*$.

Define $S = \bigcup_{i=1}^{l}\{w_1^i, w_2^i, \cdots, w_{s_i}^i\}$ as the set consisting of all source nodes that can send mass through 4-cycles. Let $S_1 = \bigcup_{t_i\in T_2}\{w_1^i, w_2^i, \cdots, w_{s_i}^i\}$ and $S_2 = \{w \in S | w \notin S_1\}$. Owing to Properties 1 and 2, we derive $\sum_{r=1}^{s_i} \pi'(w_r^i, t_i) = m(t_i)$ for $\forall t_i \in T_1$, $\sum_{r=1}^{s_i} \pi'(w_r^i, t_i) < m(t_i)$ for $\forall t_i \in T_2$, and $\sum_{t \in \mathcal{N}(w) \cap (T_1 \cup T_2)} \pi'(w,t) = m_x(w)$ for $\forall w \in S_1$. Note that $m(t_i) = m_y(t_i) - \mathbb{1}_{t_i \in \mathcal{N}(x)} \cdot m_x(t_i)$ is the upper bound of the mass that passes through 4-cycles and can be received by $t_i$, and $m_x(w)$ is the upper bound of the mass that can be sent by $w$. Thus, the total mass of $\pi'$ moved for distance 1 and passing through 4-cycles is transported from source nodes $w \in S_1 \cup S_2$ to destination nodes $t \in T_1 \cup T_2$. We divide the total mass into $a = \sum_{w \in S_2 \wedge t \in T_1} \pi'(w,t)$, $b = \sum_{w \in S_1 \wedge t \in T_1} \pi'(w,t)$, and $c = \sum_{w \in S_1 \wedge t \in T_2} \pi'(w,t)$. Based on Properties 1 and 2, we can derive that $a + b = \sum_{t_i \in T_1} m(t_i)$ and $b + c = \sum_{w \in S_1} m_x(w)$, where $\sum_{t_i \in T_1} m(t_i)$ and $\sum_{w \in S_1} m_x(w)$ are constants.

Because $a + b \geq \sum_{w \in S_2 \wedge t \in T_1} \pi^*(w,t) + \sum_{w \in S_1 \wedge t \in T_1} \pi^*(w,t)$, according to the assumption of $M_1(\pi^*) - M_1(\pi') > 0$, we can derive $c < \sum_{w \in S_1 \wedge t \in T_2} \pi^*(w,t)$, namely, $c$ can be increased. We set $\delta' < \sum_{w \in S_1 \wedge t \in T_2} \pi^*(w,t) - c$, and increase $c \leftarrow c + \delta'$. Because $b + c = \sum_{w \in S_1} m_x(w)$ is a constant, $b \leftarrow b - \delta'$. Because $a + b = \sum_{t_i \in T_1} m(t_i)$ is a constant, $a \leftarrow a + \delta'$. Thus, $M_1(\pi') = a + b + c + m_y(x) + m_x(y)$ increases by $\delta'$. Based on Eqs. (11) and (21), $M_2(\pi') + M_3(\pi')$ inevitably decreases by $\delta'$. Assuming $\delta'$ is a small positive probability mass, we can use the method in Fig. 5 to update the values of $a$, $b$ and $c$, generating a new feasible transport plan $\pi''$.

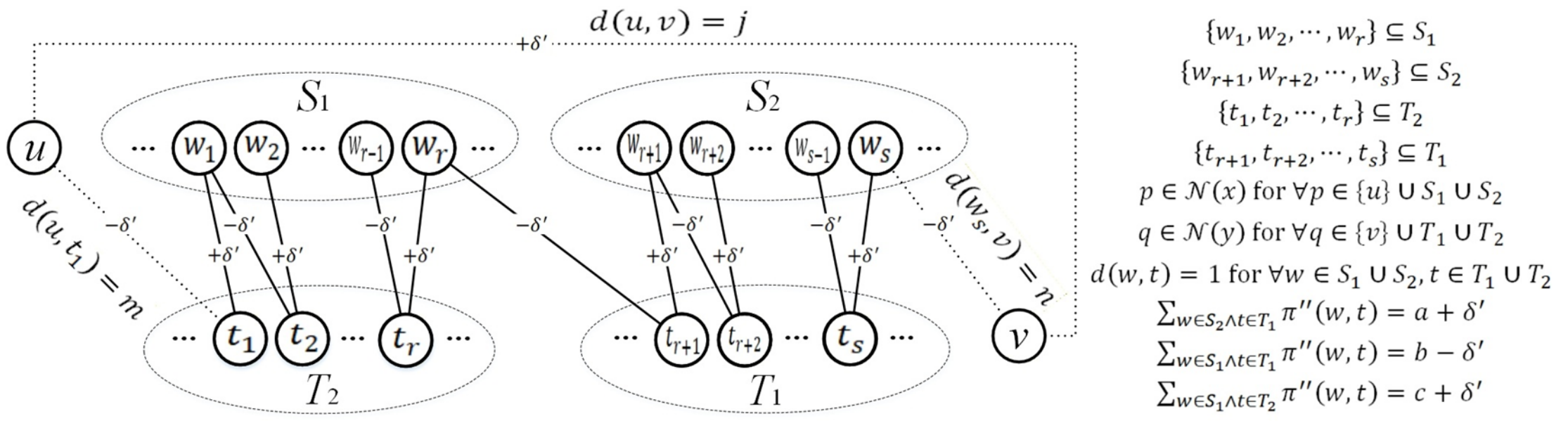

**Figure 5.** Construction of $\pi''$ for updating $c \leftarrow c + \delta'$, $b \leftarrow b - \delta'$ and $a \leftarrow a + \delta'$: $\pi''(w_i, t_i) = \pi'(w_i, t_i) + \delta'$ and $\pi''(w_i, t_{i+1}) = \pi'(w_i, t_{i+1}) - \delta'$ for $i = 1, 2, \cdots, s-1$, $\pi''(w_s, t_s) = \pi'(w_s, t_s) + \delta'$, $\pi''(w_s, v) = \pi'(w_s, v) - \delta'$, $\pi''(u, t_1) = \pi'(u, t_1) - \delta'$, $\pi''(u,v) = \pi'(u,v) + \delta'$, and $\pi''(p,q) = \pi'(p,q)$ otherwise.

Based on the constraints of $S_1$, $S_2$, $T_1$ and $T_2$ on the optimal transport plan $\pi'$, in order to increase $c$, there must exist four nodes $u \in \mathcal{N}(x) \wedge u \notin S_1 \cup S_2$, $t_1 \in T_2$, $v \in \mathcal{N}(y) \wedge v \notin T_1 \cup T_2$ and $w_s \in S_2$ satisfying $\pi'(u, t_1) > 0$ and $\pi'(w_s, v) > 0$, and there must exist a series of nodes $w_1$, $w_2, \cdots, w_r \in S_1$, $t_2, t_3, \cdots, t_r \in T_2$, $w_{r+1}, w_{r+2}, \cdots, w_{s-1} \in S_2$, and $t_{r+1}, t_{r+2}, \cdots, t_s \in T_1$ to achieve $c \leftarrow c + \delta'$, $b \leftarrow b - \delta'$ and $a \leftarrow a + \delta'$, where $\delta' = \min[\pi'(u,t_1), \pi'(w_s,v), \pi'(w_i,t_{i+1})]$, $i = 1, 2, \cdots, s-1$. Based on Eq. (8), the transport plan $\pi''$ constructed in Fig. 5 is feasible. Assuming $d(u,t_1) = m$, $d(w_s, v) = n$ and $d(u,v) = j$, we can derive $C(\pi'') - C(\pi') = (1 + j - m - n) \cdot \delta'$.

If $u = y \wedge v = x$, then $m = n = j = 1$. Thus, $C(\pi'') = C(\pi') \wedge \|M(\pi'')\| = \|M(\pi')\|$, that is, $\pi''$ is the optimal transport plan satisfying Eq. (14). However, $\pi''(y,x) = \pi'(y,x) + \delta' > 0$, which is contradictory (see the analysis in Category 1). If $u = y \wedge v \neq x$, then $n \geq 2 \wedge m = j = 1$. We can derive $C(\pi'') - C(\pi') = (1 - n) \cdot \delta' < 0$ that contradicts Eq. (14). Similarly, $u \neq y \wedge v = x$ also leads to a contradiction. Thus, $u \neq y \wedge v \neq x$, namely, $m, n, j \in \{2,3\}$.

When $m = n = 2 \wedge j = 3$, we can derive that $C(\pi'') = C(\pi') = \min_{\pi \in \Omega} C(\pi)$. Based on the construction process of $\pi''$ in Fig. 5, we derive that $M_0(\pi'') = M_0(\pi')$, $M_1(\pi'') = M_1(\pi') + \delta'$, $M_2(\pi'') = M_2(\pi') - 2\delta'$, and $M_3(\pi'') = M_3(\pi') + \delta'$. Thus, $\|M(\pi'')\| > \|M(\pi')\|$ based on Eq. (12), which contradicts Eq. (21). For other values of $m, n, j \in \{2,3\}$, $C(\pi'') - C(\pi') = (1 + j - m - n) \cdot \delta' < 0$ that contradicts Eq. (14). Therefore, $k = 1$ is contradictory.

When $k = 2$, $M_i(\pi^*) = M_i(\pi')$ for $i = 0,1$, and $\delta = M_2(\pi^*) - M_2(\pi') > 0$. Based on Eq. (17), $C(\pi^*) - C(\pi') = M_3(\pi^*) - M_3(\pi')$. Eq. (11) indicates that $M_3(\pi^*) = 1 - \sum_{i=0}^{2} M_i(\pi^*)$ and $M_3(\pi') = 1 - \sum_{i=0}^{2} M_i(\pi')$. Because $1 - \sum_{i=0}^{2} M_i(\pi^*) < 1 - \sum_{i=0}^{2} M_i(\pi')$, we can derive $C(\pi^*) - C(\pi') < 0$, namely, $C(\pi^*) < C(\pi')$ that contradicts Eq. (14).

In conclusion, $\pi^* = \pi'$ that is the optimal transport plan.

Theorem 1 provides a strict principle to approximate the exact ORC.

# 4 Curvature in cycle overlap mode (CCOM)

## 4.1 Construction of a transport plan

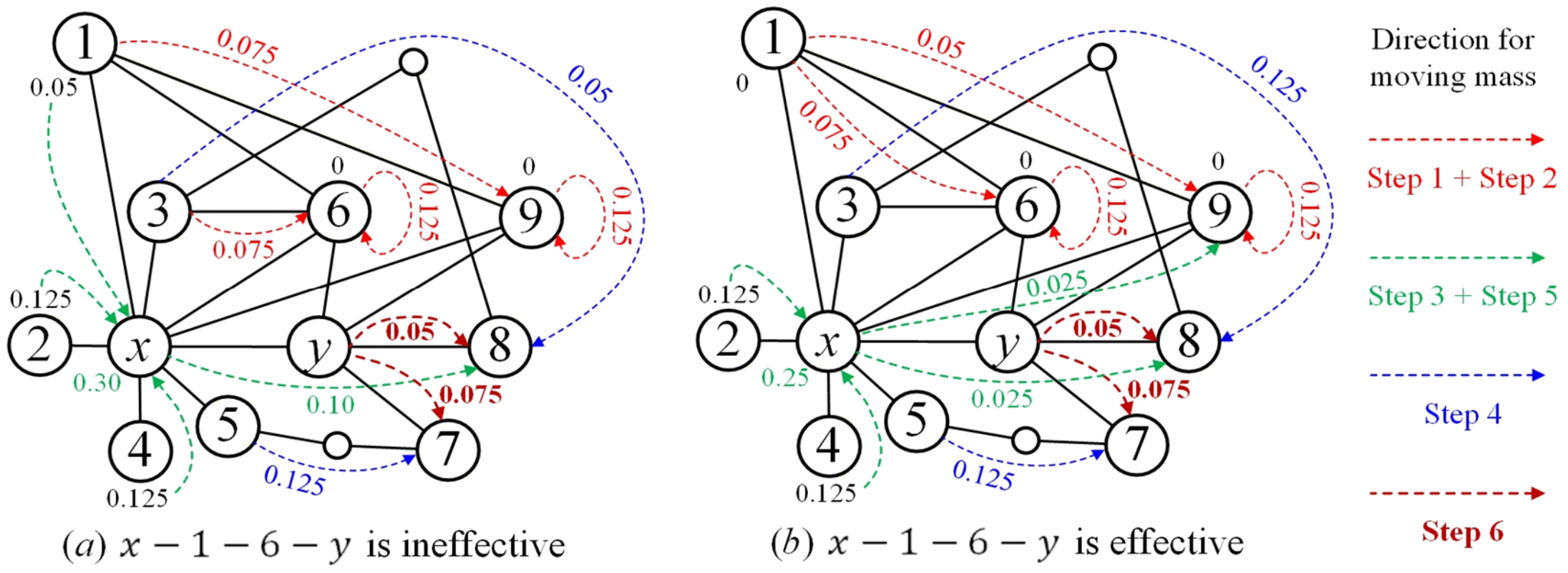

**Figure 6.** Construction of $\pi'$ to approximate $\pi^* = \text{argmax}_{\pi\in\Omega}\|M(\pi)\|$ under the constraints of $m_x(p) = 0.125, p \in \{1,2,3,4,5,6,9,y\}$ and $m_y(q) = 0.200, q \in \{6,7,8,9,x\}$. (*a*) CCOM that uses a greedy strategy to sort the cycles at Steps 2 and 4. (*b*) Inappropriate 4-cycle sorting in Step 2.

We assume that the nodes in a cycle are not duplicated and Eq. (22) holds.

To approximate $\pi^* = \text{argmax}_{\pi\in\Omega}\|M(\pi)\|$ that is an optimal transport plan proven by Theorem 1, we construct a feasible transport plan $\pi'$ as follows:

**Step 1.** Move the mass through 3-cycles to the greatest extent, namely, construct $\pi'(w,w) = \min[m_x(w), m_y(w)] = 1/d_x$ for each 3-cycle $x - w - y$, such that $M_0(\pi') = M_0(\pi^*)$.

**Step 2.** Move the remaining mass on node $u$ to node $v$ to the greatest extent for each 4-cycle $x - u - v - y$ where $u \neq v \wedge u,v \notin \{x,y\}$.

**Step 3.** Move the remaining mass on each node $u$, which cannot be moved through 5-cycles, to node $x$, where $u \in \mathcal{N}(x) \wedge u \neq y$. If the sum of the moved mass is less than $m_y(x)$, fill in the missing mass with the remaining mass on node $v$ that can be moved through 5-cycles, where $v \in \mathcal{N}(x) \wedge v \neq y$. Otherwise, temporarily store the excess mass beyond $m_y(x)$ on node $x$.

**Step 4.** Move the remaining mass on node $u$ to node $v$ to the greatest extent for each 5-cycle $x - u - w - v - y$ where $u \neq w \wedge u \neq v \wedge w \neq v \wedge u,w,v \notin \{x,y\}$.

**Step 5.** Under the constraint $\sum_{p\in\mathcal{N}(x)} \pi(p,q) = m_y(q)$, move the excess mass on node $x$ in Step 3 to the common neighbors of $x,y$ and other neighbors of $y$ in sequence.

**Step 6.** Move the mass $m_x(y) = 1/d_x$ on node $y$ to the neighbors of $y$ with missing mass.

Based on the correlation between distance and 3,4,5-cycles, as shown in Fig. 4(*a*), Steps 1 to 6 are effective to maximize $\|M(\pi')\|$. Since 3-cycles are independent of each other, Step 1 ensures that $M_0(\pi') = M_0(\pi^*)$. However, the overlapping phenomenon of 4,5-cycles is prominent in scale-free graphs [15]. Therefore, the sorting of 4-cycles in Step 2 and the sorting of 5-cycles in Step 4 are strongly related to the difference between $\pi'$ and $\pi^*$.

The 4-cycles in Fig. 6($a$) are sorted as $x-3-6-y, x-1-9-y, x-1-6-y$, and those in Fig. 6($b$) are sorted as $x-1-6-y, x-1-9-y, x-3-6-y$. Under the constraint of Eq. (8), each 4-cycle moves the remaining mass to the greatest extent, but the values of $M_1(\pi')$ in Fig. 6($a$) and Fig. 6($b$) are 0.475 and 0.450, respectively. Thus, we design a greedy strategy to sort the cycles with low time complexity. Define $\widetilde{\mathcal{N}_x}$ and $\widetilde{\mathcal{N}_y}$ as the sets of source nodes and destination nodes that may move mass through 4,5-cycles, respectively. Formally,

$$\begin{aligned}\widetilde{\mathcal{N}_x} &= \{u \mid u \in \mathcal{N}(x) \wedge u \notin \mathcal{N}(y) \wedge u \neq y\} \\ \widetilde{\mathcal{N}_y} &= \{v \mid v \in \mathcal{N}(y) \wedge v \neq x\}\end{aligned} \tag{27}$$

If $u \in \mathcal{N}(x) \wedge u \in \mathcal{N}(y)$, Step 1 constructs $\pi'(u,u) = 1/d_x$, i.e., the mass $m_x(u) = 1/d_x$ has all been moved for distance 0. Thus, the node $u$ is removed from $\widetilde{\mathcal{N}_x}$.

Define $D_{out}^k(u)$ for $\forall u \in \widetilde{\mathcal{N}_x}, k \in \{4,5\}$ as the set of destination nodes that can receive the mass through a $k$-cycle from $u$, and define $D_{in}^k(v)$ for $\forall v \in \widetilde{\mathcal{N}_y}, k \in \{3,4,5\}$ as the set of source nodes that can send the mass through a $k$-cycle to $v$. As shown in Fig. 6, $D_{out}^4(1) = \{6,9\}, D_{out}^5(1) = \emptyset$, $D_{out}^4(3) = \{6\}$, $D_{out}^5(3) = \{8\}$, $D_{in}^3(6) = \{6\}$, $D_{in}^4(6) = \{1,3\}$, $D_{in}^5(6) = \emptyset$, $D_{in}^3(9) = \{9\}$, $D_{in}^4(9) = \{1\}$, and $D_{in}^5(9) = \emptyset$. We respectively sort the source nodes of the 4-cycles in Step 2 and the source nodes of the 5-cycles in Step 4 by

$$S_{src} \leftarrow \text{Sort}\left\{u \mid u \in \widetilde{\mathcal{N}_x} \wedge \left|D_{out}^k(u)\right| > 0\right\} \text{ by } \left|D_{out}^k(u)\right| \text{ in ascending order} \tag{28}$$

where $|\cdot|$ is the cardinality of a set, $k = 4$ for Step 2, and $k = 5$ for Step 4.

We access the source nodes in sequence according to the order in $S_{src}$. For a source node $u$ being accessed in Step 2, we sort the destination nodes in $D_{out}^4(u)$ by

$$S_{tgt} \leftarrow \text{Sort } v \in D_{out}^4(u) \text{ by } \sum_{k\in\{3,4,5\}} \left|D_{in}^k(v)\right| \text{ in ascending order} \tag{29}$$

For a source node $u$ being accessed in Step 4, we sort the destination nodes in $D_{out}^5(u)$ by

$$S_{tgt} \leftarrow \text{Sort } v \in D_{out}^5(u) \text{ by } \left|D_{in}^5(v)\right| \text{ in ascending order} \tag{30}$$

We access the destination nodes in sequence according to the order in $S_{tgt}$. Fig. 6($a$) shows the result of our algorithm CCOM that adopts the greedy sorting strategy in Eqs. (28) to (30). The $k$-cycles with low values of $\left|D_{out}^k(u)\right|$ and $\left|D_{in}^k(v)\right|$ exhibit weak overlap issues, because the number of the cycles related to $u$ and $v$ is small. Our greedy strategy that prioritizes handling cycles with weak overlap can alleviate cycle competition, which helps increase $\|M(\pi')\|$.

## 4.2 Formula for CCOM

Let $F_k$ be the sum of the probability mass moved by $\pi'$ through $k$-cycles, where $k \in \{3,4,5\}$. Formally, using $\widetilde{\mathcal{N}}_x$ and $\widetilde{\mathcal{N}_y}$ in Eq. (27), we define

$$F_3 = \sum_{w\in\mathcal{N}(x)\wedge w\in\mathcal{N}(y)} \pi'(w,w) \tag{31}$$

$$F_4 = \sum_{u\in\widetilde{\mathcal{N}_x}\wedge v\in\widetilde{\mathcal{N}_y}\wedge(u,v)\in\mathcal{E}} \pi'(u,v) \tag{32}$$

$$F_5 = \sum_{u\in\widetilde{\mathcal{N}_x}\wedge v\in\widetilde{\mathcal{N}_y}\wedge(u,v)\notin\mathcal{E}\wedge\exists w\in\mathcal{N}(u)\cap\mathcal{N}(v)\wedge w\notin\{x,y\}} \pi'(u,v) \tag{33}$$

where $\mathcal{E}$ is the edge set of graph $G$, and $\pi'$ is the transport plan constructed by CCOM.

Based on Eqs. (2), (3) and (10), $\kappa_{xy} = 1 - W_1(m_x, m_y)$ and $C(\pi') = \sum_{i=0}^{3} i \cdot M_i(\pi')$ can be used to approximate $W_1(m_x, m_y)$. Thus, we can derive

$$\kappa_{xy} \approx 1 - C(\pi') = 1 - \sum_{i \in \{0,1,2,3\}} i \cdot M_i(\pi') \tag{34}$$

Define △, ■ and ⬟ represent the number of 3-cycles, 4-cycles and 5-cycles including edge $(x, y)$, respectively. In Section 4.1, Step 1 ensures that $F_3 = \triangle/d_x$. Specifically,

$$M_0(\pi') = F_3 = \triangle/d_x \tag{35}$$

Step 3 first moves the remaining mass on each node $u$, which cannot be moved through 5-cycles, to node $x$ for distance 1, where $u \in \mathcal{N}(x) \land u \neq y$, and then checks if the sum of the moved mass is less than $m_y(x)$. Define $A$ as the sum of the moved mass. For example, $A = 0.05 + 0.125 + 0.125 = 0.30$ in Fig. 6($a$), and $A = 0.125 + 0.125 = 0.25$ in Fig. 6($b$).

**Note that, the greedy sorting in Eqs. (28) and (30) may result in a part of mass not being able to be moved through 5-cycles at Step 4, which is also included in the sum $A$.**

When $A < m_y(x)$, Step 3 fills in the missing mass $m_y(x) - A$ on node $x$ with the remaining mass on node $v$ that can be moved through 5-cycles, where $v \in \mathcal{N}(x) \land v \neq y$. Define $B$ as the sum of the remaining mass. If $B \leq m_y(x) - A$, the mass $B$ is totally moved to $x$ for distance 1. Otherwise, a part of $B$ is moved by Step 4 for distance 2. When $A \geq m_y(x)$, the mass $B$ is totally moved by Step 4 for distance 2, namely $B = F_5$. For example, $B = F_5 = 0.05 + 0.125 = 0.175$ in Fig. 6($a$), and $B = F_5 = 0.125 + 0.125 = 0.25$ in Fig. 6($b$).

When $A > m_y(x)$, Step 5 moves the excess mass $A - m_y(x)$ on node $x$ in Step 3 to the common neighbors of $x, y$ and other neighbors of $y$ in sequence. If the excess mass is moved to the common neighbors $w$ where $w \in \mathcal{N}(x) \land w \in \mathcal{N}(y)$, the distance moved to $w$ is 2. For example, $d(u, 9) = 2$ for $u \in \{2,4\}$ in Fig. 6($b$). Otherwise, the excess mass is moved to other neighbors of $y$, corresponding to a distance of 3. For example, $d(u, 8) = 3$ for $u \in \{2,4\}$ in Fig. 6($b$).

Define $t$ as the sum of the missing mass on the common neighbors of $x$ and $y$ after Step 4 is executed and before Step 5 is executed. For example, $t = 0.025$ in Fig. 6($b$).

As is well known, both Step 2 and Step 6 move the mass for distance 1.

Because $\sum_{p \in \mathcal{N}(x) \land q \in \mathcal{N}(y)} \pi(p, q) = 1$, by excluding the moved mass in Step 1 ($F_3$), Step 2 ($F_4$), Step 6 ($1/d_x$) and the mass $B$, we can confirm that

$$A = \begin{cases} 1 - F_3 - F_4 - 1/d_x - B & \text{if } A < m_y(x) \\ 1 - F_3 - F_4 - 1/d_x - F_5 & \text{if } A \geq m_y(x) \end{cases} \tag{36}$$

Note that, $B = F_5$ when $A \geq m_y(x)$, $F_5$ is the mass moved by Step 4, and the mass moved by Step 5 is included in the mass $A$.

Based on the analysis of distance and mass in Steps 1 to 6 mentioned above, we derive the formula for CCOM by discussing $A$, $B$ and $t$ in the following cases:

**Case 1.** $(A \geq m_y(x) = 1/d_y) \land (A - 1/d_y \geq t) \land (t \geq 0)$

In Case 1, Step 1 moves the mass $F_3$ for distance 0, Step 2 moves the mass $F_4$ for distance 1, Step 3 moves the mass $1/d_y$ for distance 1, Step 4 moves the mass $F_5$ for distance 2, Step 5 moves the mass $t$ for distance 2 and the mass $A - 1/d_y - t$ for distance 3, and Step 6 moves the mass $1/d_x$ for distance 1. Based on Eq. (10) and Eqs. (34) to (36),

$$\kappa_{xy} \approx -2 + \frac{2}{d_x} + \frac{2}{d_y} + \frac{3}{d_x} \cdot \triangle + 2F_4 + F_5 + t \tag{37}$$

**Case 2.** $(A \geq 1/d_y) \land (A - 1/d_y < t) \land (t > 0)$

In Case 2, Step 1 moves the mass $F_3$ for distance 0, Step 2 moves the mass $F_4$ for distance 1, Step 3 moves the mass $1/d_y$ for distance 1, Step 4 moves the mass $F_5$ for distance 2, Step 5 moves the mass $A - 1/d_y$ for distance 2, and Step 6 moves the mass $1/d_x$ for distance 1. Thus,

$$\kappa_{xy} \approx -1 + \frac{1}{d_x} + \frac{1}{d_y} + \frac{2}{d_x} \cdot \triangle + F_4 \tag{38}$$

**Case 3.** $(A < 1/d_y) \wedge (B \leq 1/d_y - A)$

In Case 3, Step 1 moves the mass $F_3$ for distance 0, Step 2 moves the mass $F_4$ for distance 1, Step 3 moves the mass $A$ and the mass $B$ for distance 1, and Step 6 moves the mass $1/d_x$ for distance 1. There is no mass moved by Step 4 and Step 5. Thus,

$$\kappa_{xy} \approx \frac{1}{d_x} \cdot \triangle \tag{39}$$

**Case 4.** $(A < 1/d_y) \wedge (B > 1/d_y - A)$

In Case 4, Step 1 moves the mass $F_3$ for distance 0, Step 2 moves the mass $F_4$ for distance 1, Step 3 moves the mass $A$ and the mass $1/d_y - A$ for distance 1, Step 4 moves the mass $B - 1/d_y + A$ for distance 2, and Step 6 moves the mass $1/d_x$ for distance 1.

There is no mass moved by Step 5. Thus,

$$\kappa_{xy} \approx -1 + \frac{1}{d_x} + \frac{1}{d_y} + \frac{2}{d_x} \cdot \triangle + F_4 \tag{40}$$

In summary, for $d_x \geq d_y$ in Eq. (22), we can derive a unified formula as follows:

$$\kappa_{xy} \approx -\left(1 - \frac{1}{d_x} - \frac{1}{d_y} - \frac{1}{d_x} \cdot \triangle - F_4 - F_5 - t\right)_+ - \left(1 - \frac{1}{d_x} - \frac{1}{d_y} - \frac{1}{d_x} \cdot \triangle - F_4\right)_+ + \frac{1}{d_x} \cdot \triangle \tag{41}$$

where $(\cdot)_+$ is defined as $\max[\cdot, 0]$. Note that Eq. (41) is equivalent to Eqs. (37) to (40). Taking Case 1 as an example: the condition of Case 1 is $(A \geq m_y(x) = 1/d_y) \wedge (A - 1/d_y \geq t) \wedge (t \geq 0)$. Based on Eqs. (35) and (36), $A \geq 1/d_y \Longrightarrow 1 - 1/d_x - 1/d_y - \triangle/d_x - F_4 \geq F_5 \geq 0$, and $A - 1/d_y \geq t \Longrightarrow 1 - 1/d_x - 1/d_y - \triangle/d_x - F_4 - F_5 - t \geq 0$. Thus, Eq. (41) is transformed into $\kappa_{xy} \approx -(1 - 1/d_x - 1/d_y - \triangle/d_x - F_4 - F_5 - t) - (1 - 1/d_x - 1/d_y - \triangle/d_x - F_4) + \triangle/d_x = -2 + 2/d_x + 2/d_y + 3\,\triangle/d_x + 2F_4 + F_5 + t$ that is consistent with Eq. (37). Using the similar method, we can confirm that Eq. (41) holds for Case 2, Case 3 and Case 4.

When $d_x < d_y$, we swap $x$ and $y$ in Eq. (41), namely,

$$\kappa_{xy} \approx -\left(1 - \frac{1}{d_y} - \frac{1}{d_x} - \frac{1}{d_y} \cdot \triangle - F_4 - F_5 - t\right)_+ - \left(1 - \frac{1}{d_y} - \frac{1}{d_x} - \frac{1}{d_y} \cdot \triangle - F_4\right)_+ + \frac{1}{d_y} \cdot \triangle \tag{42}$$

In general, the formula for CCOM in the cycle overlap mode can be expressed as

$$\begin{aligned}\kappa_{xy} \approx & -\left(1 - \frac{1}{d_x} - \frac{1}{d_y} - \frac{\triangle}{\max[d_x, d_y]} - F_4 - F_5 - t\right)_+ \\ & -\left(1 - \frac{1}{d_x} - \frac{1}{d_y} - \frac{\triangle}{\max[d_x, d_y]} - F_4\right)_+ + \frac{\triangle}{\max[d_x, d_y]}\end{aligned} \tag{43}$$

According to supplementary materials at *https://doi.org/10.48550/arXiv.2606.03317*, the cycle independent mode, i.e., Eq. (7), is a particular form of the cycle overlap mode.

The maximum mass moved by a $k$-cycle including $(x, y)$, $k = 3,4,5$, is $\min[m_x(p), m_y(q)] = 1/\max[d_x, d_y]$. Thus, we define $N_{eff,k} = F_k/\min[m_x(p), m_y(q)] = F_k \cdot \max[d_x, d_y]$ as the effective number of the $k$-cycles, and transform Eq (43) into

$$\begin{aligned}\kappa_{xy} \approx & -\left(1 - \frac{1}{d_x} - \frac{1}{d_y} - \frac{\triangle + N_{eff,4} + N_{eff,5}}{\max[d_x, d_y]} - t\right)_+ \\ & -\left(1 - \frac{1}{d_x} - \frac{1}{d_y} - \frac{\triangle + N_{eff,4}}{\max[d_x, d_y]}\right)_+ + \frac{\triangle}{\max[d_x, d_y]}\end{aligned} \tag{44}$$

## 4.3 Pseudo code and flowchart

When $d_x \geq d_y$, $N_{eff,3} = \triangle$, $N_{eff,4} = F_4 \cdot d_x$ and $N_{eff,5} = F_5 \cdot d_x$. The calculations for $N_{eff,k}$, $k = 3,4,5$, occur in Steps 1, 2 and 4 of Section 4.1, respectively, and rely on $D_{out}^k(u)$ for $\forall u \in \widetilde{\mathcal{N}}_x$, $D_{in}^k(v)$ for $\forall v \in \widetilde{\mathcal{N}}_y$, and the greedy sorting strategy for the $k$-cycles. Note that all the concepts and symbols have been defined in Sections 4.1 and 4.2.

We design a cycle enumeration algorithm with pruning strategy to obtain $D_{out}^k(u)$ and $D_{in}^k(v)$. For edge $(x, y)$ in graph $G = (\mathcal{V}, \mathcal{E})$, if $u \in \mathcal{N}(x) \cap \mathcal{N}(y)$, we update $D_{in}^3(u) \leftarrow D_{in}^3(u) \cup \{u\}$, and obtain $N_{eff,3} = |\mathcal{N}(x) \cap \mathcal{N}(y)|$ where $|\cdot|$ denotes the cardinality of a set. For a node pair $u$ and $v$ where $u \in \widetilde{\mathcal{N}}_x = \{p | p \in \mathcal{N}(x) \wedge p \notin \mathcal{N}(y) \wedge p \neq y\}$ and $v \in \widetilde{\mathcal{N}}_y = \{q | q \in \mathcal{N}(y) \wedge q \neq x\}$, if $(u, v) \in \mathcal{E}$, we update $D_{out}^4(u) \leftarrow D_{out}^4(u) \cup \{v\}$ and $D_{in}^4(v) \leftarrow D_{in}^4(v) \cup \{u\}$, and discard the enumeration of all the ineffective 5-cycles $x - u - w - v - y$, because the cost of moving mass from $u$ to $v$ for distance 1 is lower. If $(u, v) \notin \mathcal{E} \wedge \exists w \in \mathcal{N}(u) \cap \mathcal{N}(v) \wedge w \notin \{x, y\}$, we update $D_{out}^5(u) \leftarrow D_{out}^5(u) \cup \{v\}$ and $D_{in}^5(v) \leftarrow D_{in}^5(v) \cup \{u\}$. The cycle enumeration algorithm includes two pruning operations: one is to remove the nodes in $\mathcal{N}(x) \cap \mathcal{N}(y)$ from $\widetilde{\mathcal{N}}_x$, the other is to discard the enumeration of the ineffective 5-cycles.

**Algorithm 1:** Main Procedure for CCOM

**Input:** Graph $G = (\mathcal{V}, \mathcal{E})$.
**Output:** Graph $G_w$ with approximated Ricci curvatures.

Initialize a weighted graph $G_W = (\mathcal{V}, \mathcal{E}, W) \leftarrow G$, where the weight $W$ on edge $(x, y) \in \mathcal{E}$ is used to store the edge curvature $\kappa_{xy}$ and is initialized to 0.
**For each** edge $e = (x, y)$ in $G$ **do**
  $d_x \leftarrow |\mathcal{N}(x)|$, $d_y \leftarrow |\mathcal{N}(y)|$.
  **If** $d_x < d_y$ **then**
    Swap $x$ and $y$, and swap $d_x$ and $d_y$.
  **End If**
  $\mathcal{N}_{com}, D_{out}^k$ for $k \in \{4,5\}, D_{in}^k$ for $k \in \{3,4,5\} \leftarrow \text{CycleEnumeration}(G, x, y)$.  //Algorithm 2
  $\kappa_{xy} \leftarrow \text{CurvatureCalculation}(\mathcal{N}_{com}, D_{out}^k, D_{in}^k, d_x, d_y, x, y)$.  //Algorithm 3
  $W(x, y) \leftarrow \kappa_{xy}$.
**End For**

**Algorithm 2:** Cycle Enumeration

**Input:** Graph $G = (\mathcal{V}, \mathcal{E})$, Edge $e = (x, y)$.
**Output:** Common neighbor set $\mathcal{N}_{com}$, Source-destination maps $D_{out}^k$ for $k \in \{4,5\}$ and $D_{in}^k$ for $k \in \{3,4,5\}$.

$\mathcal{N}_{com} \leftarrow \mathcal{N}(x) \cap \mathcal{N}(y)$.
$\widetilde{\mathcal{N}}_x \leftarrow \{u | u \in \mathcal{N}(x) \wedge u \notin \mathcal{N}(y) \wedge u \neq y\}$, $\widetilde{\mathcal{N}}_y \leftarrow \{v | v \in \mathcal{N}(y) \wedge v \neq x\}$.
Initialize $D_{out}^k(u) \leftarrow \emptyset$ for $\forall u \in \widetilde{\mathcal{N}}_x \wedge k \in \{4,5\}$, and $D_{in}^k(v) \leftarrow \emptyset$ for $\forall v \in \widetilde{\mathcal{N}}_y \wedge k \in \{4,5\}$.
**For each** $v \in \mathcal{N}_{com}$ **do** $D_{in}^3(v) \leftarrow \{v\}$ **End For**.
**For each** $u \in \widetilde{\mathcal{N}}_x$ **do**
  **For each** $v \in \widetilde{\mathcal{N}}_y$ **do**
    **If** $(u, v) \in \mathcal{E}$ **then**
      $D_{out}^4(u) \leftarrow D_{out}^4(u) \cup \{v\}$, $D_{in}^4(v) \leftarrow D_{in}^4(v) \cup \{u\}$.
      **Continue**
    **End If**
    **If** there is a node $z \in \mathcal{N}(u) \cap \mathcal{N}(v) \wedge z \notin \{x, y\}$ **then**
      $D_{out}^5(u) \leftarrow D_{out}^5(u) \cup \{v\}$, $D_{in}^5(v) \leftarrow D_{in}^5(v) \cup \{u\}$.
    **End If**
**End For**

We respectively define $s(u)$ for each source node $u \in \widetilde{\mathcal{N}}_x$ and $d(v)$ for each destination node $v \in \widetilde{\mathcal{N}}_y$ as the upper limits of probability mass that can be sent and received when they are being accessed by Steps 1 to 6 of Section 4.1. Based on the greedy sorting in Eqs. (28) to (30), when a $k$-cycle, i.e., $x-u-v-y$ or $x-u-w-v-y$, is being accessed and $d_x \geq d_y$, we update $N_{eff,k} \leftarrow N_{eff,k} + \delta \cdot d_x$ where $k \in \{4,5\}$ and $\delta = \min[s(u), d(v)]$, $s(u) \leftarrow s(u) - \delta$, and $d(v) \leftarrow d(v) - \delta$. Note that $\delta$ is the maximum mass that can be moved by the $k$-cycle. The pseudo code and flowchart of CCOM are listed in Algorithms 1 to 3 and shown in Fig. 7, respectively.

---

**Algorithm 3:** Curvature Calculation

---

**Input:** Edge $e = (x, y)$, Common neighbors $\mathcal{N}_{com}$, Source-destination maps $D^k_{out}, D^k_{in}$, Degrees $d_x, d_y$.
**Output:** Approximated Ricci curvature $\kappa_{xy}$.

$N_{eff,3} \leftarrow |\mathcal{N}_{com}|$, $N_{eff,4} \leftarrow 0$, $N_{eff,5} \leftarrow 0$.
$s(u) \leftarrow 1/d_x$ for $\forall u \in dom(D^4_{out}) \cup dom(D^5_{out})$, where $dom(D^k_{out}) = \left\{u \middle| |D^k_{out}(u)| > 0\right\}$ for $k = 4,5$.
$d(v) \leftarrow (1/d_y - 1/d_x)$ for $\forall v \in \mathcal{N}_{com}$.
$d(v) \leftarrow 1/d_y$ for $\forall v \notin \mathcal{N}_{com} \wedge v \in dom(D_{in}) = \{v \| D^4_{in}(v)| > 0 \vee |D^5_{in}(v)| > 0\}$.
**For each** $k \in \{4,5\}$ **do**
  $S_{src} \leftarrow$ Sort $u \in dom(D^k_{out})$ by $|D^k_{out}(u)|$ in ascending order (↑).
  **For each** $u \in S_{src}$ **do**
    $S_{tgt} \leftarrow$ Sort $v \in D^k_{out}(u)$ by $\sum_{i\in\{3,4,5\}} |D^i_{in}(v)|$ in ascending order (↑) for $k = 4$.
    $S_{tgt} \leftarrow$ Sort $v \in D^k_{out}(u)$ by $|D^5_{in}(v)|$ in ascending order (↑) for $k = 5$.
    **For each** $v \in S_{tgt}$ **do**
      $\delta \leftarrow \min[s(u), d(v)]$.
      Update $s(u) \leftarrow s(u) - \delta$, $d(v) \leftarrow d(v) - \delta$, $N_{eff,k} \leftarrow N_{eff,k} + \delta \cdot d_x$.
      **If** $s(u) = 0$ **then break End If**
    **End For**
  **End For**
**End For**
$A \leftarrow 1 - 1/d_x - (N_{eff,3} + N_{eff,4} + N_{eff,5})/d_x$.
**If** $A < 1/d_y$ **then** #Cases 3 and 4 in Section 4.2
  Update $N_{eff,5} \leftarrow N_{eff,5} - \min[(1/d_y - A) \cdot d_x, N_{eff,5}]$.
**End If**
Calculate $\kappa_{xy}$ using $N_{eff,4}$, $N_{eff,5}$, $\triangle = N_{eff,3}$, and $t = \sum_{v\in\mathcal{N}_{com}} d(v)$ via Eq. (44).

---

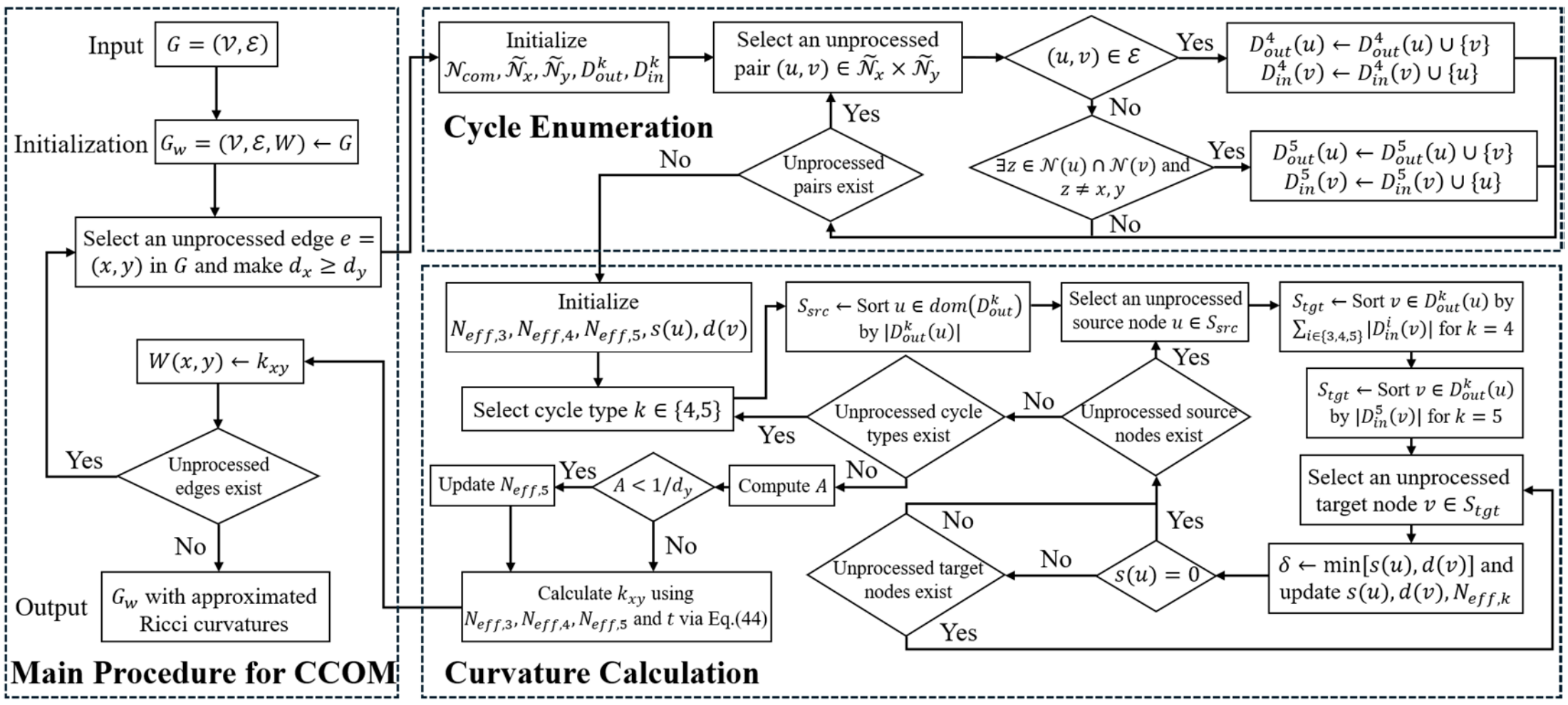

**Figure 7.** Overall flowchart of CCOM.

# 5 Time complexity and error bound

## 5.1 Time complexity of CCOM

We analyze the time consumption on edge $(x, y)$ for calculating $\kappa_{xy}$ with $d_x \geq d_y$. The time is mainly spent on Algorithms 2 and 3. In Algorithm 2, the 3-cycle enumeration $\mathcal{N}_{com} \leftarrow \mathcal{N}(x) \cap \mathcal{N}(y)$ adopts a HashSet operation with a complexity of $O(d_y)$, and the enumeration of 4,5-cycles relies on two loops of $u$ in $\widetilde{\mathcal{N}}_x$ and $v$ in $\widetilde{\mathcal{N}}_y$, where $|\widetilde{\mathcal{N}}_x| = d_x - |\mathcal{N}_{com}| - 1$ and $|\widetilde{\mathcal{N}}_y| = d_y - 1$. In each iteration of the loops, the 4-cycle enumeration needs to determine whether $(u, v) \in \mathcal{E}$ is true, with a complexity of $O(1)$, while the 5-cycle enumeration needs to determine whether a node $z \in \mathcal{N}(u) \cap \mathcal{N}(v) \wedge z \notin \{x, y\}$ exists, with a complexity of $O(\min[d_u, d_v])$. Thus, the time complexity of Algorithm 2 can be expressed as

$$T_{Enumeration} = O(d_y) + \sum_{u \in \widetilde{\mathcal{N}}_x} \sum_{v \in \widetilde{\mathcal{N}}_y} \{O(1) + \mathbb{1}_{(u,v) \notin \mathcal{E}} \cdot O(\min[d_u, d_v])\} \tag{45}$$

where $\mathbb{1}_{(u,v) \notin \mathcal{E}} = 1$ if $(u, v) \notin \mathcal{E}$, otherwise $0$.

In Algorithm 3, the time is mainly spent on the sorting operations in lines 6 and 8 (or 9). Then, the time complexity of Algorithm 3 can be expressed as

$$T_{Curvature} = O\left(\left|dom\left(D_{out}^k\right)\right| \cdot \log\left(\left|dom\left(D_{out}^k\right)\right|\right)\right) + O\left(\left|D_{out}^k(u)\right| \cdot \log\left(\left|D_{out}^k(u)\right|\right)\right) \tag{46}$$

where $|dom(D_{out}^k)| \le |\widetilde{\mathcal{N}}_x| < d_x$ and $|D_{out}^k(u)| \le |\widetilde{\mathcal{N}}_y| < d_y$.

Scale-free graphs have core-periphery structures. Specifically, a large number of low-degree nodes fall in the periphery [15], which helps reduce the complexity of Eqs. (45) and (46).

## 5.2 Error bound of CCOM

The event of moving probability mass through 4-cycles only occurs in Step 2 of Section 4.1. Define $U = \{u | u \in \widetilde{\mathcal{N}}_x \wedge \exists v \in \widetilde{\mathcal{N}}_y, (u, v) \in \mathcal{E}\}$ as the set of source nodes that can send mass through 4-cycles $x - u - v - y$, define $V = \{v | v \in \widetilde{\mathcal{N}}_y \wedge \exists u \in \widetilde{\mathcal{N}}_x, (u, v) \in \mathcal{E}\}$ as the set of destination nodes that can receive mass through the 4-cycles, and define $E = \{(u, v) \in \mathcal{E} | u \in U \wedge v \in V\}$ as the set of edges that connect $u \in U$ and $v \in V$ through the 4-cycles. Then, we construct a bipartite graph $H = (U, V, E)$. After Step 1 of Section 4.1 is executed, define $s(u)$ for each source node $u \in U$ and $d(v)$ for each destination node $v \in V$ as the upper limits of probability mass that can be sent and received, respectively. Then, we transform Step 2 of Section 4.1 into a mass allocation problem on the bipartite graph $H = (U, V, E)$ with node mass $s(u)$ and $d(v)$.

Let $\pi^*$ and $\pi'$ be the optimal transport plan and an approximation transport plan obtained by our approach CCOM, respectively. Since $M_0(\pi^*) = M_0(\pi')$ after Step 1 of Section 4.1, the bipartite graph $H = (U, V, E)$ with node mass $s(u)$ and $d(v)$ provides the same initial state for $\pi^*$ and $\pi'$ before executing step 2. Define $f_{\mathrm{Opt}} = \{f_{uv}^*\}_{(u,v) \in E}$ and $f_{\mathrm{CCOM}} = \{f_{uv}'\}_{(u,v) \in E}$ as the set of mass moved by $\pi^*$ and $\pi'$ on the bipartite graph $H$, respectively.

**Theorem 2**. $Q_4^*/2 < Q_4' \le Q_4^*$, where $Q_4^* = \sum_{(u,v) \in E} f_{uv}^*$ and $Q_4' = \sum_{(u,v) \in E} f_{uv}'$.

**Proof.** At Step 2 of Section 4.1, CCOM moves the mass $f_{uv}' = \min[s_r(u), d_r(v)]$ for each 4-cycle $x - u - v - y$ that is being accessed, where $s_r(u)$ and $d_r(v)$ respectively indicate the upper limits of mass that can be sent and received in the current state, and updates $s_r(u) \leftarrow s_r(u) - f_{uv}'$, and $d_r(v) \leftarrow d_r(v) - f_{uv}'$. Thus, when Step 2 is completed by $\pi'$, $s_r(u) = 0 \vee d_r(v) = 0$, namely, $s(u) = \sum_{v:(u,v) \in E} f_{uv}' \vee d(v) = \sum_{u:(u,v) \in E} f_{uv}'$, for each $(u, v) \in E$ in $H$. Define $S_U = \{u \in U | s(u) = \sum_{v:(u,v) \in E} f_{uv}'\}$, $S_V = \{v \in V | d(v) = \sum_{u:(u,v) \in E} f_{uv}'\}$, Then,

$$\sum_{u \in S_U} s(u) + \sum_{v \in S_V} d(v) = \sum_{(u,v) \in E} f_{uv}' \cdot \left(\mathbb{1}_{u \in S_U} + \mathbb{1}_{v \in S_V}\right) \tag{47}$$

where $\mathbb{1}_{u \in S_U} = 1$ if $u \in S_U$, and $\mathbb{1}_{v \in S_V} = 1$ if $v \in S_V$, otherwise $0$.

Because $s(u) = \sum_{v:(u,v)\in E} f'_{uv} \vee d(v) = \sum_{u:(u,v)\in E} f'_{uv}$ for each $(u,v) \in E$ in $H$,

$$\begin{aligned} Q_4^* &= \sum_{(u,v)\in E} f_{uv}^* \le \sum_{(u,v)\in E} f_{uv}^* \cdot \left(\mathbb{1}_{u\in S_U} + \mathbb{1}_{v\in S_V}\right) \\ &= \sum_{u\in S_U} \sum_{v:(u,v)\in E} f_{uv}^* + \sum_{v\in S_V} \sum_{u:(u,v)\in E} f_{uv}^* \end{aligned} \tag{48}$$

Because $\sum_{v:(u,v)\in E} f_{uv}^* \le s(u) \wedge \sum_{u:(u,v)\in E} f_{uv}^* \le d(v)$, based on Eq. (48),

$$Q_4^* \le \sum_{u\in S_U} s(u) + \sum_{v\in S_V} d(v) \tag{49}$$

Because $\mathbb{1}_{u\in S_U} + \mathbb{1}_{v\in S_V} \le 2$ for each $(u,v) \in E$, based on Eqs. (47) and (49),

$$Q_4^* \le 2\sum_{(u,v)\in E} f'_{uv} = 2Q'_4 \tag{50}$$

Note that, in Eq. (50), $Q_4^* = 2Q'_4$ if and only if $\mathbb{1}_{u\in S_U} + \mathbb{1}_{v\in S_V} = 2$ for $\forall(u,v) \in E$. In other words, $u \in S_U \wedge v \in S_V$ holds for $\forall(u,v) \in E$, namely, $\sum_{u\in U} s(u) = \sum_{(u,v)\in E} f'_{uv} = \sum_{v\in V} d(v)$. That is a conflict, because $Q'_4 = \sum_{(u,v)\in E} f'_{uv} = \sum_{u\in U} s(u) \ge \sum_{(u,v)\in E} f_{uv}^* = Q_4^*$.

Thus, $Q_4^*/2 < Q'_4$. As shown in Fig. 4($a$), distance 1 corresponds to moving through a 4-cycle, moving from neighbors of $x$ to $x$, and moving from $y$ to neighbors of $y$. Because Steps 3 and 6 of Section 4.1 maximize the mass of moving from neighbors of $x$ to $x$, and the mass of moving from $y$ to neighbors of $y$, respectively, and $M_1(\pi') \le M_1(\pi^*)$ where $\pi^* = \text{argmax}_{\pi\in\Omega}\|M(\pi)\|$, we can derive that $Q'_4 \le Q_4^*$. In summary, $Q_4^*/2 < Q'_4 \le Q_4^*$. The proof is completed.

Let $\kappa_{xy}^*$ and $\kappa'_{xy}$ be the exact and our approximate curvatures on edge $(x,y)$, respectively. According to the proof of Theorem 2, $M_0(\pi^*) = M_0(\pi')$ and $M_1(\pi^*) - M_1(\pi') = Q_4^* - Q'_4$. Thus, based on Eqs. (2), (3), (10), (11) and (13), we can derive

$$\kappa_{xy}^* - \kappa'_{xy} = C(\pi') - C(\pi^*) = 2(Q_4^* - Q'_4) + [M_2(\pi^*) - M_2(\pi')] \ge 0 \tag{51}$$

As shown in Fig. 4($a$), distance 2 corresponds to moving through a 5-cycle and moving from a neighbor of $x$ to a common neighbor of $x$ and $y$. Thus, $0 \le M_2(\pi^*), M_2(\pi') \le \pentagon/d_x + \triangle\cdot(1/d_y - 1/d_x)$, and $M_2(\pi^*) - M_2(\pi') < \pentagon/d_x + \triangle\cdot\left(1/d_y - 1/d_x\right)$. Based on Theorem 2, $2(Q_4^* - Q'_4) < Q_4^* \le \blacksquare/d_x$, where $\triangle$, $\blacksquare$ and $\pentagon$ denote the number of 3-cycles, 4-cycles and 5-cycles including edge $(x,y)$, respectively. In summary, an error boundary of CCOM can be expressed as

$$0 \le \kappa_{xy}^* - \kappa'_{xy} < \blacksquare/d_x + \pentagon/d_x + \triangle\cdot\left(1/d_y - 1/d_x\right) \tag{52}$$

In addition, the optimal transport principle proved by Theorem 1 in Section 3 provides a more rigorous theoretical basis for error controllability of CCOM, because the design of CCOM in Section 4.1 is based on the optimal transport principle. The superiority of our approach in error controllability will be further experimentally verified in Section 6.

# 6 Experiments

This section validates the effectiveness of CCOM on curvature approximation and community detection, and compares it with baseline approaches including MWPM [19], LoCur [14], ALU [25], LRC [10], FRC [26] and BFC [13]. Because the direct link between LRC (FRC,BFC) and ORC is less straightforward [10], we only compare the approximation error of CCOM with MWPM, LoCur and ALU. The exact ORC was calculated in parallel [6] using the linear programming model in Eq. (3), and the community detection was executed within the same curvature-based framework in Fig. 2. Experiments were conducted on a laptop computer equipped with a 2.20 GHz 14th Gen Intel(R) Core(TM) i7-14650HX CPU and 32 GB of RAM, and were implemented in Python 3.

The analysis of MWPM [19] is based on an ORC that differs slightly from Eq. (3), but it can be extended to the ORC in Section 2.1 (see Appendix A). For a fair comparison, the exact ORC approximated by CCOM and baselines remains consistent with the definition in Section 2.1.

The real-world datasets used for comparison were collected from Stanford Network Analysis Project [35], PyTorch Geometric [36], Network Repository [37], Mark Newman's personal website [38] and NetworkX [39]. All the collected datasets were converted into undirected, unweighted and simple graphs, and their statistics and descriptions are listed in Table 1.

**Table 1.** Dataset statistics and descriptions. The self-loops, multi-edges, edge-directions, and edge-weights of the graphs have been removed.

| **Datasets** | **# Nodes** | **# Edges** | **Average Degree** | **Number of 3-cycles** | **Descriptions** |
|---|---|---|---|---|---|
| Florentine families [39] | 15 | 20 | 2.6667 | 3 | Social network focusing on the marriage and business ties between elite families |
| Karate [39] | 34 | 78 | 4.5882 | 45 | Social network of friendships between the 34 members of a karate club |
| DD [36] | 37 | 94 | 5.0811 | 86 | One graph extracted from DD |
| Dolphin [38] | 62 | 159 | 5.1290 | 95 | Social network of frequent associations between 62 dolphins |
| Les miserable [39] | 77 | 254 | 6.5974 | 467 | Coappearance network of characters in the novel Les Miserables |
| Polbooks [38] | 105 | 441 | 8.4000 | 560 | Network of 105 books on American politics |
| Football [38] | 115 | 613 | 10.6608 | 810 | Network of American football games |
| REDDIT-5K [36] | 521 | 585 | 2.2457 | 14 | One graph extracted from REDDIT-5K |
| Cora [36] | 2,708 | 5,278 | 3.8981 | 1,630 | Citation network |
| Citeseer [36] | 3,327 | 4,552 | 2.7364 | 1,167 | Citation network |
| PubMed [36] | 19,717 | 44,324 | 4.4960 | 12,520 | Citation network |
| com-LiveJournal [35] | 39,982 | 300,892 | 15.0514 | 1,749,002 | LiveJournal social network and ground-truth communities |
| com-DBLP [35] | 317,080 | 1,049,866 | 6.6221 | 2,224,385 | DBLP collaboration network and ground-truth communities |
| com-Amazon [35] | 334,863 | 925,872 | 5.5299 | 667,129 | Amazon product co-purchasing network and ground-truth communities |
| ca-IMDB [37] | 896,308 | 3,782,447 | 8.4401 | 4,358 | IMDB movie/actor network |
| roadNet-PA [35] | 1,088,092 | 1,541,898 | 2.8341 | 67,150 | Pennsylvania road network |
| cit-Patents [35] | 3,774,768 | 16,518,947 | 8.7523 | 7,515,023 | Patent citation network |

In Table 1, DD and REDDIT-5K [36] are datasets for graph classification. We choose one graph from each of them with low error of LoCur for curvature approximation comparison. Similar to [40], we retain the first 3,750 communities in the com-LiveJournal dataset [35] to reduce the influence of low-quality communities in the community detection task.

### 6.1 Error comparison

Let $\{\hat{\kappa}_{xy} | (x,y) \in \mathcal{E}\}$ be the set of edge curvatures calculated by an approximation approach, and $\{\kappa_{xy} | (x,y) \in \mathcal{E}\}$ be the set of the exact ORC on graph $G = (\mathcal{V}, \mathcal{E})$. Then, we evaluate the error from two perspectives. One is to compare the mean values of the approximate and exact curvatures, namely, $\widehat{\text{Mean}}$ and $\text{Mean}$ in Eq. (53), and the other is to calculate the mean of absolute error and the mean of relative error, namely, $\text{MAE}$ and $\text{MRE}$ in Eq. (54).

$$\widehat{\text{Mean}} = \frac{1}{|\mathcal{E}|} \sum_{(x,y)\in\mathcal{E}} \hat{\kappa}_{xy}, \quad \text{Mean} = \frac{1}{|\mathcal{E}|} \sum_{(x,y)\in\mathcal{E}} \kappa_{xy} \tag{53}$$

$$\text{MAE} = \frac{1}{|\mathcal{E}|} \sum_{(x,y)\in\mathcal{E}} |\hat{\kappa}_{xy} - \kappa_{xy}|, \quad \text{MRE} = \frac{1}{|\tilde{\mathcal{E}}|} \sum_{(x,y)\in\tilde{\mathcal{E}}} \left| \frac{\hat{\kappa}_{xy} - \kappa_{xy}}{\kappa_{xy}} \right| \times 100\% \tag{54}$$

where $\tilde{\mathcal{E}} = \{(x,y) \in \mathcal{E} | \kappa_{xy} \neq 0\}$. Because the exact ORC $\{\kappa_{xy} | (x,y) \in \mathcal{E}\}$ can only be calculated on small-scale graphs, we choose the datasets in Table 1 with no more than 20,000 nodes for error comparison, and the results are listed in Table 2.

**Table 2.** Comparison of calculation errors for the curvature approximation approaches.

| Datasets | CCOM | | MWPM | | LoCur | | ALU | | Exact ORC |
|---|---|---|---|---|---|---|---|---|---|
| | $\widehat{\text{Mean}}$ | MAE / MRE | $\widehat{\text{Mean}}$ | MAE / MRE | $\widehat{\text{Mean}}$ | MAE / MRE | $\widehat{\text{Mean}}$ | MAE / MRE | Mean |
| Florentine families | **-0.0833** | **1.76e-16** | -0.2250 | 0.1417 | -0.2583 | 0.1750 | -0.0708 | 0.0625 | **-0.0833** |
| | | **< 0.01%** | | 83.90% | | 81.75% | | 25.79% | |
| Karate | **0.0055** | **0.0012** | -0.1519 | 0.1587 | -0.3938 | 0.4005 | -0.1105 | 0.1582 | **0.0067** |
| | | **1.71%** | | 128.82% | | 423.23% | | 173.84% | |
| DD | **0.2479** | **0.0003** | 0.0622 | 0.1859 | 0.1779 | 0.0702 | 0.3208 | 0.0771 | **0.2481** |
| | | **0.20%** | | 107.48% | | 78.43% | | 55.23% | |
| Dolphin | **-0.0577** | **0.0034** | -0.3172 | 0.2630 | -0.5182 | 0.4639 | -0.1542 | 0.1492 | **-0.0542** |
| | | **10.96%** | | 273.65% | | 449.18% | | 139.35% | |
| Les miserables | **0.2087** | **0.0007** | -0.0067 | 0.2161 | 0.0248 | 0.1847 | 0.2268 | 0.0954 | **0.2095** |
| | | **1.20%** | | 202.13% | | 176.19% | | 147.30% | |
| Polbooks | **0.0289** | **0.0053** | -0.2508 | 0.2850 | -0.5631 | 0.5973 | -0.1589 | 0.2239 | **0.0342** |
| | | **5.60%** | | 302.11% | | 748.67% | | 284.46% | |
| Football | **0.0437** | **0.0047** | -0.0217 | 0.0701 | -0.5326 | 0.5810 | -0.0879 | 0.1786 | **0.0484** |
| | | **2.37%** | | 31.88% | | 215.26% | | 66.27% | |
| REDDIT-5K | **-0.2952** | **0.0001** | -0.3263 | 0.0312 | -0.3283 | 0.0332 | -0.1594 | 0.1395 | **-0.2951** |
| | | **0.03%** | | 20.59% | | 16.85% | | 45.12% | |
| Cora | **-0.3993** | **0.0013** | -0.5386 | 0.1407 | -0.6287 | 0.2307 | -0.2609 | 0.2165 | **-0.3980** |
| | | **0.58%** | | 77.67% | | 115.34% | | 58.56% | |
| Citeseer | **-0.2436** | **0.0019** | -0.3777 | 0.1360 | -0.4736 | 0.2319 | -0.1897 | 0.1597 | **-0.2417** |
| | | **1.29%** | | 95.18% | | 159.22% | | 73.74% | |
| PubMed | **-0.6166** | **0.0048** | -0.6914 | 0.0797 | -0.9976 | 0.3858 | -0.4820 | 0.2375 | **-0.6118** |
| | | **2.53%** | | 47.46% | | 189.68% | | 80.39% | |

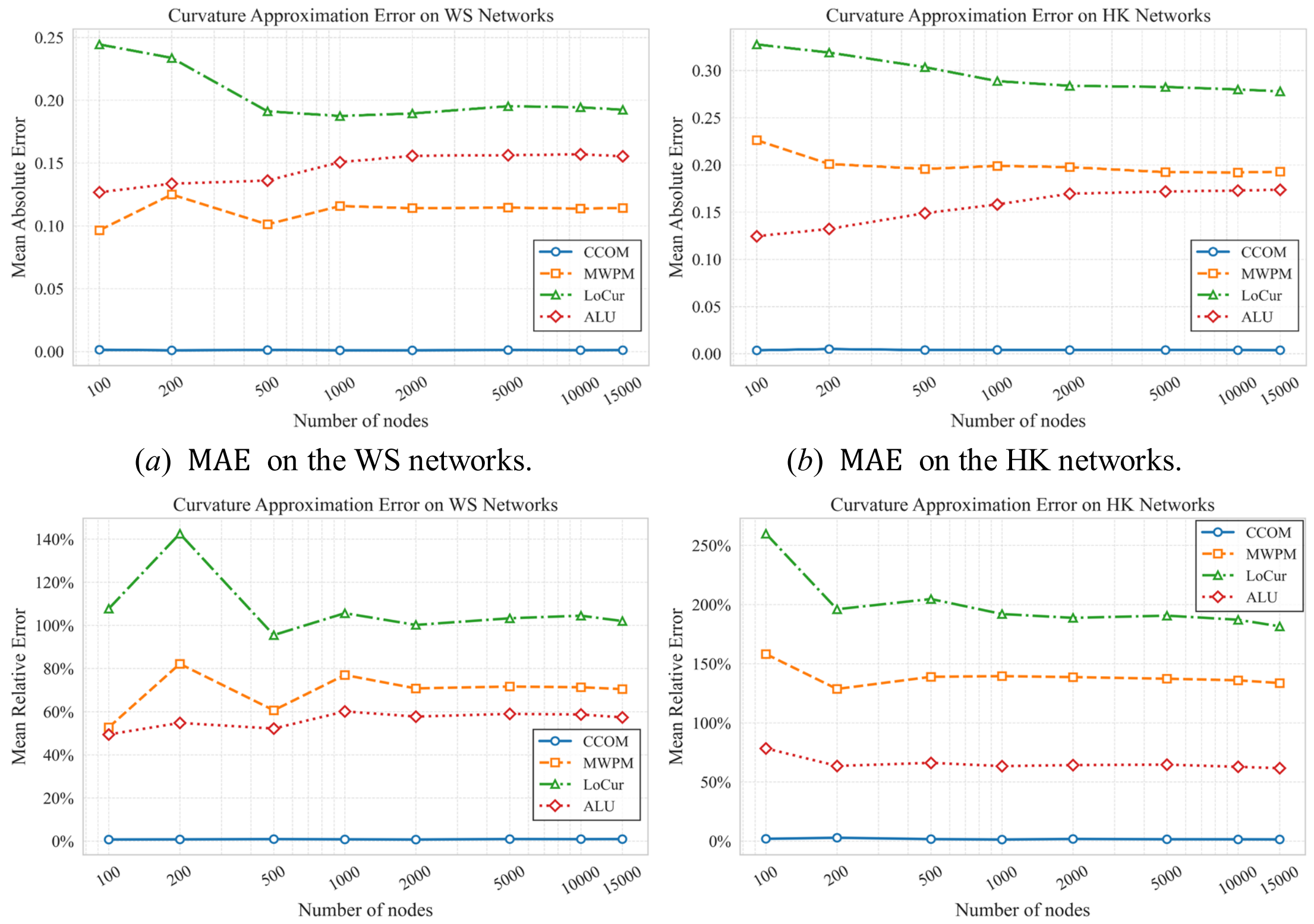

(*a*) MAE on the WS networks. (*b*) MAE on the HK networks.

(*c*) MRE on the WS networks. (*d*) MRE on the HK networks.

**Figure 8.** Comparison of MAE and MRE on simulated networks with increasing scale. In the python experiments, we remove the edges $(x,y)$ with $|\kappa_{xy}| \leq 10^{-12}$ from $\tilde{\mathcal{E}}$ in Eq. (54).

As shown in Table 2, the $\widehat{\mathrm{Mean}}$ values of CCOM are closest to the Mean values of the exact ORC, and the MAE and MRE values of CCOM are significantly lower than those of baselines. Thus, CCOM performs superior in terms of accuracy. In addition, we choose the Watts-Strogatz (WS) [41] and Holme-Kim (HK) [42] generators to simulate small-world and scale-free networks, respectively, and capture MAE and MRE on these networks with increasing scale. The WS generator starts with a ring of $n$ nodes, each connected to its $k$ neighbors, and then reconnects edges with probability $p$. We set $k = 6$ and $p = 0.1$ for the WS networks. At each step of the HK generator, one node with $m$ edges is preferentially attached to existing high-degree nodes, and a parameter $p$ is used to adjust the number of triangles. We set $m = 3$ and $p = 1.0$ for the HK networks. As shown in Fig. 8, the errors of CCOM on small-world and scale-free networks are very small, and can be controlled as the scale of the networks increases.

### 6.2 Time and memory comparison

We compare CCOM with baselines in terms of running time and memory usage, measured in seconds (s) and megabytes (MB). The exact ORC was calculated using open-source parallel code [6]: *https://github.com/saibalmars/GraphRicciCurvature*, while CCOM and baseline approximation approaches were calculated using single threading.

**Table 3.** Comparison of running time (seconds) and memory usage (megabytes) for the exact and approximation approaches for ORC calculation.

| Datasets | CCOM | | MWPM | | LoCur | | ALU | | Exact ORC | |
|---|---|---|---|---|---|---|---|---|---|---|
| | Time (s) | Memory (MB) | Time (s) | Memory (MB) | Time (s) | Memory (MB) | Time (s) | Memory (MB) | Time (s) | Memory (MB) |
| Florentine families | 0.0002 | 964.5 | 0.0002 | 964.5 | 0.0001 | 964.4 | 0.0001 | 964.5 | 0.1611 | 964.6 |
| Karate | 0.0013 | 964.6 | 0.0013 | 964.5 | 0.0002 | 964.5 | 0.0002 | 964.4 | 0.1512 | 964.9 |
| DD | 0.0010 | 974.0 | 0.0010 | 974.1 | 0.0002 | 974.0 | 0.0002 | 973.9 | 0.1861 | 974.2 |
| Dolphin | 0.0027 | 964.7 | 0.0025 | 964.8 | 0.0004 | 964.7 | 0.0004 | 964.7 | 0.1680 | 965.3 |
| Les miserables | 0.0072 | 964.6 | 0.0076 | 964.8 | 0.0005 | 964.5 | 0.0006 | 964.5 | 0.1535 | 965.4 |
| Polbooks | 0.0179 | 965.0 | 0.0147 | 964.9 | 0.0008 | 964.7 | 0.0008 | 964.8 | 0.1654 | 965.4 |
| Football | 0.0176 | 965.1 | 0.0173 | 965.0 | 0.0011 | 964.8 | 0.0011 | 964.8 | 0.1747 | 965.6 |
| REDDIT-5K | 0.0076 | 968.1 | 0.0450 | 968.7 | 0.0013 | 968.1 | 0.0014 | 968.2 | 0.2212 | 973.6 |
| Cora | 0.1221 | 970.1 | 0.1551 | 970.6 | 0.0129 | 970.1 | 0.0141 | 970.1 | 0.5992 | 1189.3 |
| Citeseer | 0.0895 | 970.8 | 0.0951 | 970.6 | 0.0117 | 970.6 | 0.0135 | 970.6 | 0.7526 | 1478.2 |
| PubMed | 3.6722 | 980.1 | 3.9046 | 979.7 | 0.1391 | 979.4 | 0.1528 | 979.5 | 25.3 | 18755.6 |
| com-LiveJournal | 13.5 | 8703.6 | 24.5 | 8704.0 | 0.7284 | 8703.3 | 0.7414 | 8703.5 | OOM | OOM |
| com-DBLP | 172.1 | 1695.3 | 213.0 | 1685.6 | 4.83 | 1684.7 | 4.84 | 1685.1 | OOM | OOM |
| com-Amazon | 32.3 | 1627.6 | 35.6 | 1617.0 | 4.73 | 1616.4 | 4.98 | 1616.7 | OOM | OOM |
| ca-IMDB | 2706.9 | 3903.9 | 3442.4 | 3893.3 | 24.4 | 3872.4 | 25.6 | 3872.1 | OOM | OOM |
| roadNet-PA | 11.1 | 2512.0 | 13.7 | 2472.3 | 5.81 | 2472.0 | 5.91 | 2471.9 | OOM | OOM |
| cit-Patents | 4400.8 | 13761.5 | 4398.7 | 13632.8 | 126.1 | 13627.3 | 129.6 | 13627.5 | OOM | OOM |

Table 3 shows that the time and memory consumption of the exact ORC significantly increase on the PubMed graph with 19,717 nodes and 44,324 edges [36]. Owing to an out of memory (OOM) error, it is difficult to obtain the exact ORC on graphs with more than 20,000 nodes, whereas CCOM and baselines perform superior in terms of running time and memory usage, making ORC applicable to larger-scale graphs. Owing to the enumeration of 3,4,5-cycles, CCOM exhibits higher time consumption compared with LoCur and ALU. However, a large number of 3,4,5-cycles overlap within scale-free graphs [15,23], and are critical for reducing the computational error of ORC, as shown in Table 2 and Fig. 8. We can observe that the time consumption of CCOM is close to that of MWPM, and the memory consumption of CCOM is close to those of LoCur, ALU and MWPM.

## 6.3 Community detection

Using the same framework in Fig. 2, we compare CCOM with baselines, including LRC, FRC, BFC, LoCur, ALU and MWPM, in the application of community detection tasks. We choose the six datasets of Dolphin, Polbooks, Football, com-LiveJournal, com-DBLP and com-Amazon in Table 1 for the comparison, because the datasets provide ground-truth communities, which are necessary for the evaluation criteria of ARI, AMI and F1-score discussed in Section 2.3. Specifically, ARI and AMI [32,33] are evaluated for single-membership communities, and F1-score [34] is evaluated for mixed-membership communities.

### 6.3.1 Single-membership comparison

The inputs of the curvature-based framework in Fig. 2 include $K$ (i.e., the predefined number of communities) and $S_{min}$ (i.e., the number of nodes in the smallest community). We set $K$ and $S_{min}$ according to the ground-truth community labels for the single-membership datasets, including Dolphin, Polbooks, and Football. The comparison results of CCOM and baselines for detecting single-membership communities are listed in Tables 4 and 5, and the visualization of the datasets of Dolphin and Football are shown in Figs. 9 and 10, respectively. The results confirm the superior performance of CCOM on the single-membership community detection tasks.

**Table 4.** Community detection on graphs with single-membership communities.

| Curvatures | Dolphin | | Polbooks | | Football | |
|---|---|---|---|---|---|---|
| | ARI | AMI | ARI | AMI | ARI | AMI |
| **LRC** | 0.8721 | 0.8117 | 0.4460 | 0.4254 | 0.8543 | 0.8793 |
| **FRC** | 0.8721 | 0.8117 | 0.4322 | 0.4623 | 0.8543 | 0.8793 |
| **BFC** | 0.9348 | 0.8874 | 0.4460 | 0.4254 | 0.8543 | 0.8793 |
| **LoCur** | 0.8721 | 0.8117 | 0.4396 | 0.4508 | 0.8543 | 0.8793 |
| **ALU** | 0.8721 | 0.8117 | 0.4466 | 0.4549 | 0.8543 | 0.8793 |
| **MWPM** | 0.8721 | 0.8117 | 0.5009 | 0.4893 | 0.8543 | 0.8793 |
| **CCOM** | **1.0000** | **1.0000** | **0.6745** | **0.5652** | **0.8893** | **0.9029** |

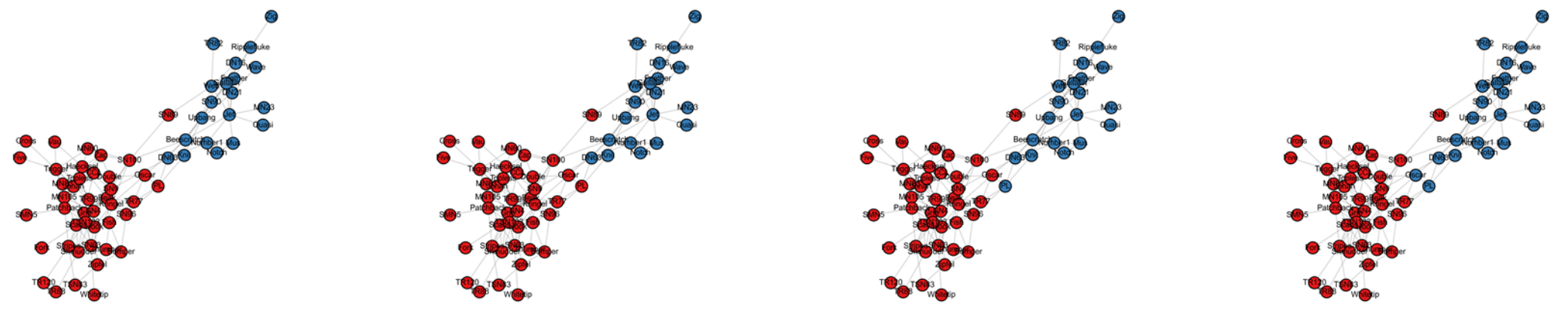

($a$) Ground Truth ($b$) CCOM ($c$) BFC ($d$) Others (FRC, LoCur, etc.)

**Figure 9.** Visualization of community detection on Dolphin with 62 nodes and 159 edges [38].

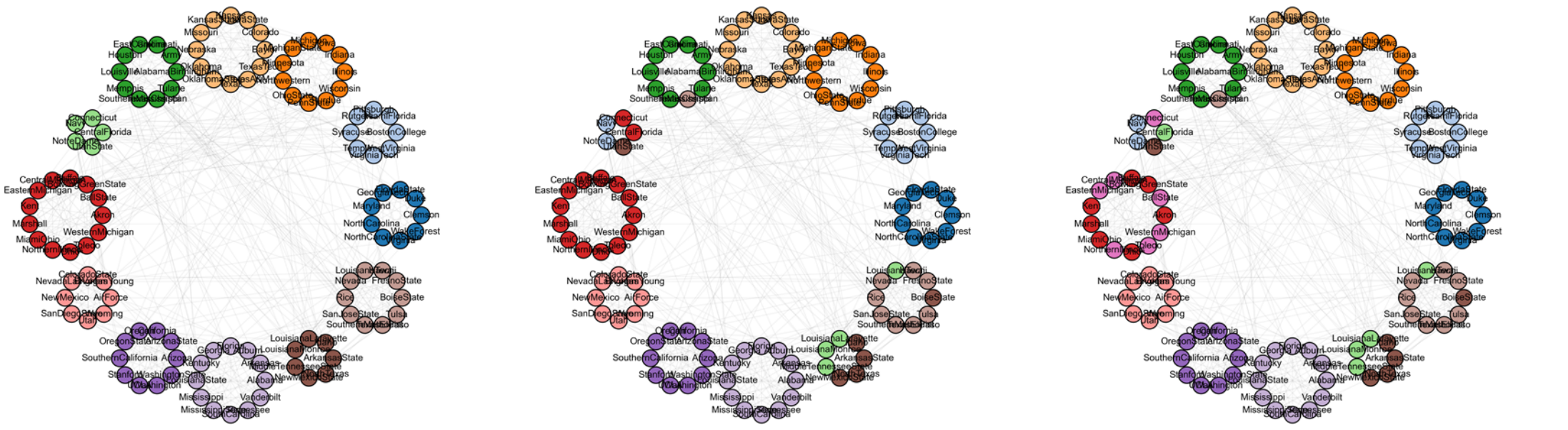

($a$) Ground Truth ($b$) CCOM ($c$) Others (FRC, BFC, LoCur, etc.)

**Figure 10.** Visualization of community detection on Football with 115 nodes and 613 edges [38].

**Table 5.** Community detection on Football before and after the preferential attachment of Fig. 2 (# Comms denotes the number of detected communities).

| Curvatures | Football (before) | | | Football (after) | | |
|---|---|---|---|---|---|---|
| | ARI | AMI | # Comms | ARI | AMI | # Comms |
| **LRC** | 0.8697 | 0.8906 | 14 | 0.8543 | 0.8793 | 13 |
| **FRC** | 0.8755 | 0.8970 | 15 | 0.8543 | 0.8793 | 13 |
| **BFC** | 0.8600 | 0.8857 | 14 | 0.8543 | 0.8793 | 13 |
| **LoCur** | 0.8755 | 0.8970 | 15 | 0.8543 | 0.8793 | 13 |
| **ALU** | 0.8600 | 0.8857 | 14 | 0.8543 | 0.8793 | 13 |
| **MWPM** | 0.8453 | 0.8687 | 16 | 0.8543 | 0.8793 | 13 |
| **CCOM** | **0.8893** | **0.9029** | **12** | **0.8893** | **0.9029** | **12** |

#### 6.3.2 Mixed-membership comparison

Three large-scale graphs, i.e., com-LiveJournal, com-DBLP and com-Amazon, are used for the mixed-membership comparison. We set the inputs $K = None$ and $S_{min} = None$ for the three graphs, and adopt the top 5,000 highest-quality ground-truth communities in the datasets of com-DBLP and com-Amazon to evaluate the F1-score. In addition, we choose the top 3,750 highest-quality ground-truth communities in the dataset of com-LiveJournal to construct a subgraph induced by the communities, and adopt the 3,750 highest-quality ground-truth communities to evaluate the F1-score of the communities that are detected in the subgraph. Table 6 shows the superior performance of CCOM on large-scale and mixed-membership community detection tasks. Because BFC [13] manipulates the adjacency matrices of input graphs, which are prone to OOM errors on large-scale graphs, it has not been applied to community detection tasks for nodes over 30,000.

**Table 6.** Comparison of F1-score for detecting mixed-membership communities.

| Curvatures | com-LiveJournal | com-DBLP | com-Amazon |
|---|---|---|---|
| | 39,982 nodes/ 300,892 edges | 317,080 nodes/ 1,049,866 edges | 334,863 nodes/ 925,872 edges |
| **LRC** | 0.8577 | 0.6445 | 0.7668 |
| **FRC** | 0.8379 | 0.6385 | 0.7695 |
| **LoCur** | 0.8538 | 0.6339 | 0.7543 |
| **ALU** | 0.8562 | 0.6434 | 0.8342 |
| **MWPM** | 0.7923 | 0.5738 | 0.6555 |
| **CCOM** | **0.8878** | **0.6473** | **0.8535** |

## 7 Conclusions

A scale-free graph consists of a dense (small) core and a sparse (large) periphery [23], and the centrum of the core, composed of a few core nodes with top highest degrees, is densely connected by the periphery. Specifically, there are numerous 3,4,5-cycles passing through the centrum nodes, which contribute to reducing the distance between low-degree nodes and give rise to the small-world property [15]. Thus, the centrum results in extensive overlapping of the 3,4,5-cycles.

This paper rigorously proves the optimal transport principle of ORC in the cycle overlap mode, and proposes an approach CCOM based on the principle to approximate the exact ORC. By designing the strategies of cycle enumeration, pruning search, and greedy sorting, CCOM greatly reduces the computational error. In addition, experiments verified that the time complexity of CCOM is close to that of MWPM, and confirmed that the memory consumption of CCOM is close to those of LoCur, ALU and MWPM. The advantage of CCOM in accuracy demonstrates its superior performance in community detection tasks. In the future, we will extend the steps for constructing a transport plan in Section 4.1 to the ORC defined by different $\alpha$ values in Eq. (1), and further explore the application of CCOM in the field of deep learning on large scale-free networks.

## CRediT authorship contribution statement:

**Zexian Zhou:** Software; Data curation; Methodology; Formal analysis; Writing - Original draft.

**Bo Jiao:** Conceptualization; Methodology; Formal analysis; Supervision; Writing - Review & Editing.

**Corresponding author:** Bo Jiao (jiaoboleetc@outlook.com)

## **Appendix A:** Extension of MWPM to the ORC defined in Section 2.1

In Section 2.1, when $\alpha = 0$, the probability measure $m_x$ at node $x$ is defined as [2]:

$$m_x(p) = \begin{cases} 1/d_x & \text{if } p \in \mathcal{N}(x) \\ 0 & \text{otherwise} \end{cases} \quad (\mathrm{A.1})$$

However, the probability measure $m_x$ of the original MWPM is defined as [19]:

$$m_x(p) = \begin{cases} 1/(d_x + 1) & \text{if } p \in \mathcal{N}(x) \cup \{x\} \\ 0 & \text{otherwise} \end{cases} \quad (\mathrm{A.2})$$

We assume that $d_x \le d_y$. Let $\mathcal{N}(x) \cup \{x\} = \{p_1, p_2, \cdots, p_{d_x+1}\}$, and $\mathcal{N}(y) \cup \{y\} = \{q_1, q_2, \cdots, q_{d_y+1}\}$. Then, the original MWPM derives two integers $a \ge 1$ and $0 \le b < d_x + 1$, satisfying $d_y + 1 = a(d_x + 1) + b$, and constructs a weighted bipartite graph $G_B = \{L, R, W\}$ as follows:

**Step 1.** Set $L = \{p_1, \cdots, p_{d_x+1}\}$ and $R = \{q_1, \cdots, q_{d_y+1}\}$.

**Step 2.** Replace each node $p_i \in L$ by $a$ nodes $p_i^1, \cdots, p_i^a$, and add $b$ additional nodes $r_1, \cdots, r_b$ into $L$, such that $|L| = |R| = d_y + 1$, where $|\cdot|$ denotes the cardinality of a set.

**Step 3.** Set weights $W(p_i^j, q_\ell) = d(p_i, q_\ell)$ for $\forall i \in \{1, \cdots, d_x + 1\}$, $j \in \{1, \cdots, a\}$, and $\ell \in \{1, \cdots, d_y + 1\}$, where $d(p_i, q_\ell)$ denotes the shortest path length between $p_i$ and $q_\ell$, and set weights $W(r_i, q_\ell) = 3$ for $\forall i \in \{1, \cdots, b\}$ and $\ell \in \{1, \cdots, d_y + 1\}$.

Then, the original MWPM uses the minimum-weight perfect matching on $G_B$ to approximate the 1-Wasserstein distance $W_1(m_x, m_y)$ [19]. Let $\mathcal{N}(x) = \{p_1, p_2, \cdots, p_{d_x}\}$ and $\mathcal{N}(y) = \{q_1, q_2 \cdots, q_{d_y}\}$. We derive two integers $a \ge 1$ and $0 \le b < d_x$, satisfying $d_y = a \cdot d_x + b$, and constructs a weighted bipartite graph $G_B' = \{L', R', W'\}$ as follows:

**Step 1.** Set $L' = \{p_1, \cdots, p_{d_x}\}$ and $R' = \{q_1, \cdots, q_{d_y}\}$.

**Step 2.** Replace each node $p_i \in L'$ by $a$ nodes $p_i^1, \cdots, p_i^a$, and add $b$ additional nodes $r_1, \cdots, r_b$ into $L'$, such that $|L'| = |R'| = d_y$, where $|\cdot|$ denotes the cardinality of a set.

**Step 3.** Set weights $W'(p_i^j, q_\ell) = d(p_i, q_\ell)$ for $\forall i \in \{1, \cdots, d_x\}$, $j \in \{1, \cdots, a\}$, and $\ell \in \{1, \cdots, d_y\}$, where $d(p_i, q_\ell)$ denotes the shortest path length between $p_i$ and $q_\ell$, and set weights $W'(r_i, q_\ell) = 3$ for $\forall i \in \{1, \cdots, b\}$ and $\ell \in \{1, \cdots, d_y\}$.

Then, based on the analysis in [19], the minimum-weight perfect matching on $G_B'$ can be used to approximate the 1-Wasserstein distance $W_1(m_x, m_y)$ defined in Section 2.1.

# Supplementary Materials

## SM.1 ORC in the cycle independent mode

### SM.1.1 Definitions and symbols

For an edge $(x,y)$ in a simple, undirected and unweighted graph $G=(\mathcal{V},\mathcal{E})$, a 3-cycle including $(x,y)$ is represented as $x-w-y$ where $w\in\mathcal{N}(x)\cap\mathcal{N}(y)$, a 4-cycle including $(x,y)$ is represented as $x-u-v-y$ where $u\in\mathcal{N}(x)\wedge v\in\mathcal{N}(y)\wedge(u,v)\in\mathcal{E}\wedge u,v\notin\{x,y\}$, a 5-cycle including $(x,y)$ is represented as $x-u-w-v-y$ where $u\in\mathcal{N}(x)\wedge v\in\mathcal{N}(y)\wedge w\in\mathcal{N}(u)\cap\mathcal{N}(v)\wedge u\neq v\wedge u,w,v\notin\{x,y\}$. We define $\triangle$, ■ and ⬟ as the number of 3-cycles, 4-cycles and 5-cycles including edge $(x,y)$, respectively, and define the following symbols:

The set of common neighbors of $x,y$ is defined as $\mathcal{N}_{com}=\mathcal{N}(x)\cap\mathcal{N}(y)$, i.e., $|\mathcal{N}_{com}|=\triangle$.

The set of 4-cycle neighbors of $x$ is defined as $\mathcal{N}_4(x)=\{u\in\mathcal{N}(x)\wedge u\neq y|\exists v\in\mathcal{N}(y)\wedge v\notin\{x,u\},(u,v)\in\mathcal{E}\}$. Similarly, $\mathcal{N}_4(y)$ is the set of 4-cycle neighbors of $y$.

The set of 5-cycle neighbors of $x$ is defined as $\mathcal{N}_5(x)=\{u\in\mathcal{N}(x)\wedge u\neq y|\exists v\in\mathcal{N}(y)\wedge v\notin\{x,u\}\wedge w\notin\{x,y\},w\in\mathcal{N}(u)\cap\mathcal{N}(v)\}$.

Similarly, $\mathcal{N}_5(y)$ is defined as the set of 5-cycle neighbors of $y$.

The set of exclusive neighbors of $x$ is defined as $\mathcal{N}_{exc}(x)=\{u\in\mathcal{N}(x)\wedge u\neq y|u\notin\mathcal{N}_{com}\wedge u\notin\mathcal{N}_4(x)\wedge u\notin\mathcal{N}_5(x)\}$. Similarly, $\mathcal{N}_{exc}(y)$ is defined as the set of exclusive neighbors of $y$.

As shown in Fig. SM-1, $\mathcal{N}_{com}=\{6\}$, $\mathcal{N}_4(x)=\{4\}$, $\mathcal{N}_4(y)=\{8\}$, $\mathcal{N}_5(x)=\{5\}$, $\mathcal{N}_5(y)=\{9\}$, $\mathcal{N}_{exc}(x)=\{1,2,3\}$, and $\mathcal{N}_{exc}(y)=\{10,11\}$.

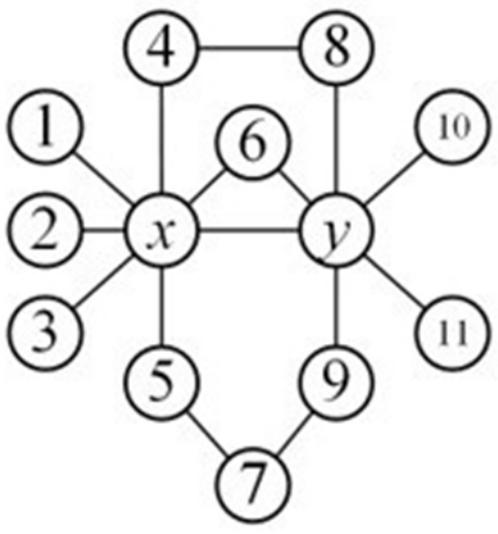

**Figure SM-1.** Cycle independent mode.

In the cycle independent mode, all 3,4,5-cycles including $(x,y)$ are mutually independent, i.e., any two of them do not include the same node except for $x$ and $y$. Therefore, we can confirm that $|\mathcal{N}_4(x)|=|\mathcal{N}_4(y)|=$ ■, $|\mathcal{N}_5(x)|=|\mathcal{N}_5(y)|=$ ⬟, $|\mathcal{N}_{exc}(x)|=d_x-1-\triangle-$ ■ $-$ ⬟ and $|\mathcal{N}_{exc}(y)|=d_y-1-\triangle-$ ■ $-$ ⬟, where $d_x$ and $d_y$ denote the degree of $x$ and $y$, respectively.

### SM.1.2 Formula for the cycle independent mode

Without loss of generality, we assume that $d_x\geq d_y$. Then, we construct particular solutions for the linear optimization problem in Eq. (3) and the Kantorovich duality in Eq. (4) to derive the upper and lower bounds of the 1-Wasserstein distance $W_1(m_x,m_y)$, respectively.

A feasible transport plan $\pi'$ is constructed using the following steps:

**Step 1.** Move the mass through 3-cycles to the greatest extent, namely, construct $\pi'(w,w)=min[m_x(w),m_y(w)]=1/d_x$ for each 3-cycle $x-w-y$.

Eqs. (2), (3), (4), (5) and (41), used in supplementary materials, are defined in the manuscript titled Ollivier-Ricci curvature in cycle overlap mode.

**Step 2.** Move the mass through 4-cycles to the greatest extent, namely, construct $\pi'(u,v) = min[m_x(u), m_y(v)] = 1/d_x$ for each 4-cycle $x-u-v-y$.

**Step 3.** Let $A' = \sum_{q\in\mathcal{N}_{exc}(y)} m_y(q) - m_x(y) = 1 - 1/d_x - 1/d_y - (\triangle + \blacksquare + \blacklozenge)/d_y$. If $A' < 0$, node $y$ has no exclusive neighbors (see Appendix SM.A for proof), therefore Step 3 is skipped. Otherwise, move the mass $m_x(y) = 1/d_x$ at node $y$ to $\mathcal{N}_{exc}(y)$.

**Step 4.** Let $B' = \sum_{p\in\mathcal{N}_{exc}(x)} m_x(p) - m_y(x) = 1 - 1/d_x - 1/d_y - (\triangle + \blacksquare + \blacklozenge)/d_x$. If $B' \geq 0$, move the mass $m_y(x) = 1/d_y$ at $\mathcal{N}_{exc}(x)$ to node $x$, otherwise move the mass $\sum_{p\in\mathcal{N}_{exc}(x)} m_x(p) = 1 - 1/d_x - (\triangle + \blacksquare + \blacklozenge)/d_x$ at $\mathcal{N}_{exc}(x)$ to node $x$.

**Step 5.** If $B' < 0$, move a part of mass at $\mathcal{N}_5(x)$ to node $x$ to fill the missing mass $m_y(x) - \sum_{p\in\mathcal{N}_{exc}(x)} m_x(p)$, and then move the mass through 5-cycles to the greatest extent. Otherwise, construct $\pi'(u,v) = min[m_x(u), m_y(v)] = 1/d_x$ for each 5-cycle $x-u-w-v-y$.

**Step 6.** Under the constraint $\sum_{p\in\mathcal{N}(x)} \pi'(p,q) = m_y(q)$, move the remaining mass at $\mathcal{N}_{exc}(x)$ to the common neighbors of $x, y$ and other neighbors of $y$ in sequence. If $A' < 0$ in Step 3, move the mass $m_x(y) = 1/d_x$ at node $y$ to the neighbors of $y$ to fill the missing mass.

The transport cost $C(\pi')$ depends on the sign of $A'$ and $B'$. $A' - B' = 1/d_x - 1/d_y \leq 0$, we therefore consider the following cases:

**Case 1.** $0 \leq A' \leq B'$

Step 1 moves the mass $\triangle/d_x$ for distance 0, Step 2 moves the mass $\blacksquare/d_x$ for distance 1, Step 3 moves the mass $1/d_x$ for distance 1, Step 4 moves the mass $1/d_y$ for distance 1, Step 5 moves the mass $\blacklozenge/d_x$ for distance 2, and Step 6 moves the mass $(1/d_y - 1/d_x)\cdot\triangle$ for distance 2, and the mass $B' - (1/d_y - 1/d_x)\cdot\triangle$ for distance 3.

Based on the optimization problem in Eq. (3),

$$W_1(m_x, m_y) \leq 3 - \frac{2}{d_x} - \frac{2}{d_y} - \left(\frac{2}{d_x} + \frac{1}{d_y}\right)\cdot\triangle - \frac{2}{d_x}\cdot\blacksquare - \frac{1}{d_x}\cdot\blacklozenge \quad \text{(SM1)}$$

Construct a 1-Lipschitz function $f$ of the Kantorovich duality in Eq. (4):

$$f(z) = \begin{cases} 0, & z \in \mathcal{N}_{exc}(y) \cup \mathcal{N}_4(y) \cup \mathcal{N}_5(y) \\ 1, & z \in \mathcal{N}_{com} \cup \{y\} \cup \mathcal{N}_4(x) \\ 2, & z \in \{x\} \cup \mathcal{N}_5(x) \\ 3, & z \in \mathcal{N}_{exc}(x) \end{cases} \quad \text{(SM2)}$$

Based on Eq. (SM2) and the Kantorovich duality in Eq. (4),

$$W_1(m_x, m_y) \geq 3 - \frac{2}{d_x} - \frac{2}{d_y} - \left(\frac{2}{d_x} + \frac{1}{d_y}\right)\cdot\triangle - \frac{2}{d_x}\cdot\blacksquare - \frac{1}{d_x}\cdot\blacklozenge \quad \text{(SM3)}$$

Owing to the upper bound in Eq. (SM1) and the lower bound in Eq. (SM3),

$$W_1(m_x, m_y) = 3 - \frac{2}{d_x} - \frac{2}{d_y} - \left(\frac{2}{d_x} + \frac{1}{d_y}\right)\cdot\triangle - \frac{2}{d_x}\cdot\blacksquare - \frac{1}{d_x}\cdot\blacklozenge \quad \text{(SM4)}$$

Based on Eq. (2), we can derive

$$\kappa_{xy} = -2 + \frac{2}{d_x} + \frac{2}{d_y} + \left(\frac{2}{d_x} + \frac{1}{d_y}\right)\cdot\triangle + \frac{2}{d_x}\cdot\blacksquare + \frac{1}{d_x}\cdot\blacklozenge \quad \text{(SM5)}$$

**Case 2.** $A' < 0 \leq B'$

$A' < 0$ indicates that node $y$ has no exclusive neighbors (see Appendix SM.A).

**Case 2.1** $A' < 0 \leq B' \wedge B' \leq (1/d_y - 1/d_x)\cdot\triangle$

Step 1 moves the mass $\triangle/d_x$ for distance 0, Step 2 moves the mass $\blacksquare/d_x$ for distance 1, Step 3 is skipped, Step 4 moves the mass $1/d_y$ for distance 1, Step 5 moves the mass $\blacklozenge/d_x$ for distance 2, and Step 6 moves the mass $B'$ for distance 2, and the mass $1/d_x$ for distance 1.

Based on the optimization problem in Eq. (3),

$$W_1(m_x, m_y) \le 2 - \frac{1}{d_x} - \frac{1}{d_y} - \frac{2}{d_x} \cdot \triangle - \frac{1}{d_x} \cdot \blacksquare \tag{SM6}$$

Construct a 1-Lipschitz function $f$ of the Kantorovich duality in Eq. (4):

$$f(z) = \begin{cases} 0, & z \in \mathcal{N}_{com} \cup \mathcal{N}_4(y) \cup \mathcal{N}_5(y) \\ 1, & z \in \{x, y\} \cup \mathcal{N}_4(x) \\ 2, & z \in \mathcal{N}_{exc}(x) \cup \mathcal{N}_5(x) \end{cases} \tag{SM7}$$

Based on Eq. (SM7) and the Kantorovich duality in Eq. (4),

$$W_1(m_x, m_y) \ge 2 - \frac{1}{d_x} - \frac{1}{d_y} - \frac{2}{d_x} \cdot \triangle - \frac{1}{d_x} \cdot \blacksquare \tag{SM8}$$

Owing to the upper bound in Eq. (SM6) and the lower bound in Eq. (SM8),

$$W_1(m_x, m_y) = 2 - \frac{1}{d_x} - \frac{1}{d_y} - \frac{2}{d_x} \cdot \triangle - \frac{1}{d_x} \cdot \blacksquare \tag{SM9}$$

Based on Eq. (2), we can derive

$$\kappa_{xy} = -1 + \frac{1}{d_x} + \frac{1}{d_y} + \frac{2}{d_x} \cdot \triangle + \frac{1}{d_x} \cdot \blacksquare \tag{SM10}$$

**Case 2.2** $A' < 0 \le B' \wedge B' > (1/d_y - 1/d_x) \cdot \triangle$

Step 1 moves the mass $\triangle/d_x$ for distance 0, Step 2 moves the mass $\blacksquare/d_x$ for distance 1, Step 3 is skipped, Step 4 moves the mass $1/d_y$ for distance 1, Step 5 moves the mass $\pentagon/d_x$ for distance 2, and Step 6 moves the mass $(1/d_y - 1/d_x) \cdot \triangle$ for distance 2, the mass $B' - (1/d_y - 1/d_x) \cdot \triangle$ for distance 3, and the mass $1/d_x$ for distance 1.

Based on the optimization problem in Eq. (3),

$$W_1(m_x, m_y) \le 3 - \frac{2}{d_x} - \frac{2}{d_y} - \left(\frac{2}{d_x} + \frac{1}{d_y}\right) \cdot \triangle - \frac{2}{d_x} \cdot \blacksquare - \frac{1}{d_x} \cdot \pentagon \tag{SM11}$$

Construct a 1-Lipschitz function $f$ of the Kantorovich duality in Eq. (4):

$$f(z) = \begin{cases} 0, & z \in \mathcal{N}_4(y) \cup \mathcal{N}_5(y) \\ 1, & z \in \mathcal{N}_{com} \cup \{y\} \cup \mathcal{N}_4(x) \\ 2, & z \in \{x\} \cup \mathcal{N}_5(x) \\ 3, & z \in \mathcal{N}_{exc}(x) \end{cases} \tag{SM12}$$

Based on Eq. (SM12) and the Kantorovich duality in Eq. (4),

$$W_1(m_x, m_y) \ge 3 - \frac{2}{d_x} - \frac{2}{d_y} - \left(\frac{2}{d_x} + \frac{1}{d_y}\right) \cdot \triangle - \frac{2}{d_x} \cdot \blacksquare - \frac{1}{d_x} \cdot \pentagon \tag{SM13}$$

Owing to the upper bound in Eq. (SM11) and the lower bound in Eq. (SM13),

$$W_1(m_x, m_y) = 3 - \frac{2}{d_x} - \frac{2}{d_y} - \left(\frac{2}{d_x} + \frac{1}{d_y}\right) \cdot \triangle - \frac{2}{d_x} \cdot \blacksquare - \frac{1}{d_x} \cdot \pentagon \tag{SM14}$$

Based on Eq. (2), we can derive

$$\kappa_{xy} = -2 + \frac{2}{d_x} + \frac{2}{d_y} + \left(\frac{2}{d_x} + \frac{1}{d_y}\right) \cdot \triangle + \frac{2}{d_x} \cdot \blacksquare + \frac{1}{d_x} \cdot \pentagon \tag{SM15}$$

**Case 3.** $A' \le B' < 0$

$A' < 0$ indicates that node $y$ has no exclusive neighbors (see Appendix SM.A).

**Case 3.1** $A' \le B' < 0 \wedge (-B' \ge \pentagon/d_x)$

Step 1 moves the mass $\triangle/d_x$ for distance 0, Step 2 moves the mass $\blacksquare/d_x$ for distance 1, Step 3 is skipped, Step 4 moves the mass $1 - 1/d_x - (\triangle + \blacksquare + \pentagon)/d_x$ for distance 1, Step 5 moves the mass $\pentagon/d_x$ for distance 1, and Step 6 moves the mass $1/d_x$ for distance 1.

Based on the optimization problem in Eq. (3),

$$W_1(m_x, m_y) \leq 1 - \frac{1}{d_x} \cdot \triangle \tag{SM16}$$

Construct a 1-Lipschitz function $f$ of the Kantorovich duality in Eq. (4):

$$f(z) = \begin{cases} 0, & z \in \mathcal{N}_{com} \cup \{x\} \cup \mathcal{N}_4(y) \cup \mathcal{N}_5(y) \\ 1, & z \in \{y\} \cup \mathcal{N}_4(x) \cup \mathcal{N}_5(x) \cup \mathcal{N}_{exc}(x) \end{cases} \tag{SM17}$$

Based on Eq. (SM17) and the Kantorovich duality in Eq. (4),

$$W_1(m_x, m_y) \geq 1 - \frac{1}{d_x} \cdot \triangle \tag{SM18}$$

Owing to the upper bound in Eq. (SM16) and the lower bound in Eq. (SM18),

$$W_1(m_x, m_y) = 1 - \frac{1}{d_x} \cdot \triangle \tag{SM19}$$

Based on Eq. (2), we can derive

$$\kappa_{xy} = \frac{1}{d_x} \cdot \triangle \tag{SM20}$$

**Case 3.2** $A' \leq B' < 0 \wedge (-B' < ⬟/d_x)$

Step 1 moves the mass $\triangle/d_x$ for distance 0, Step 2 moves the mass $■/d_x$ for distance 1, Step 3 is skipped, Step 4 moves the mass $1 - 1/d_x - (\triangle + ■ + ⬟)/d_x$ for distance 1, Step 5 moves the mass $-B'$ for distance 1, and the mass $⬟/d_x + B'$ for distance 2, and Step 6 moves the mass $1/d_x$ for distance 1. Based on the optimization problem in Eq. (3),

$$W_1(m_x, m_y) \leq 2 - \frac{1}{d_x} - \frac{1}{d_y} - \frac{2}{d_x} \cdot \triangle - \frac{1}{d_x} \cdot ■ \tag{SM21}$$

Construct a 1-Lipschitz function $f$ of the Kantorovich duality in Eq. (4):

$$f(z) = \begin{cases} 0, & z \in \mathcal{N}_4(y) \cup \mathcal{N}_5(y) \\ 1, & z \in \mathcal{N}_{com} \cup \{y\} \cup \mathcal{N}_4(x) \\ 2, & z \in \{x\} \cup \mathcal{N}_5(x) \\ 3, & z \in \mathcal{N}_{exc}(x) \end{cases} \tag{SM22}$$

Based on Eq. (SM22) and the Kantorovich duality in Eq. (4),

$$W_1(m_x, m_y) \geq 2 - \frac{1}{d_x} - \frac{1}{d_y} - \frac{2}{d_x} \cdot \triangle - \frac{1}{d_x} \cdot ■ \tag{SM23}$$

Owing to the upper bound in Eq. (SM21) and the lower bound in Eq. (SM23),

$$W_1(m_x, m_y) = 2 - \frac{1}{d_x} - \frac{1}{d_y} - \frac{2}{d_x} \cdot \triangle - \frac{1}{d_x} \cdot ■ \tag{SM24}$$

Based on Eq. (2), we can derive

$$\kappa_{xy} = -1 + \frac{1}{d_x} + \frac{1}{d_y} + \frac{2}{d_x} \cdot \triangle + \frac{1}{d_x} \cdot ■ \tag{SM25}$$

In summary, for $d_x \geq d_y$, we can derive a unified formula as follows:

$$\kappa_{xy} = -\left(1 - \frac{1}{d_x} - \frac{1}{d_y} - \frac{\triangle}{d_y} - \frac{■ + ⬟}{d_x}\right)_+ - \left(1 - \frac{1}{d_x} - \frac{1}{d_y} - \frac{\triangle + ■}{d_x}\right)_+ + \frac{\triangle}{d_x} \tag{SM26}$$

where $(\cdot)_+$ is defined as $\max[\cdot, 0]$. Note that, Eq. (SM26) is equivalent to Eqs. (SM5), (SM10), (SM15), (SM20) and (SM25). Taking Case 1 as an example: the condition of Case 1 is $0 \leq A' \leq B'$; thus, we obtain that $A' \geq 0 \Longrightarrow 1 - 1/d_x - 1/d_y - (\triangle + ■ + ⬟)/d_y \geq 0 \Longrightarrow 1 - 1/d_x - 1/d_y - \triangle/d_y - (■ + ⬟)/d_x \geq 0$ and that $B' \geq 0 \Longrightarrow 1 - 1/d_x - 1/d_y - (\triangle + ■ + ⬟)/d_x \geq 0 \Longrightarrow 1 - 1/d_x - 1/d_y - (\triangle + ■)/d_x \geq 0$, that is, Eq. (SM26) can be transformed into $\kappa_{xy} = -(1 - 1/d_x -$ $1/d_y - \triangle/d_y - (\blacksquare + \blacklozenge)/d_x) - (1 - 1/d_x - 1/d_y - (\triangle + \blacksquare)/d_x) + \triangle/d_x = -2 + 2/d_x + 2/d_y + (2/d_x + 1/d_y) \cdot \triangle + 2 \cdot \blacksquare/d_x + \blacklozenge/d_x$ that is consistent with Eq. (SM5). Using the similar method, we can confirm that Eq. (SM26) holds for Case 2 and Case 3.

When $d_x < d_y$, we swap $x$ and $y$ in Eq. (SM26), namely,

$$\kappa_{xy} = -\left(1 - \frac{1}{d_x} - \frac{1}{d_y} - \frac{\triangle}{d_x} - \frac{\blacksquare + \blacklozenge}{d_y}\right)_+ - \left(1 - \frac{1}{d_x} - \frac{1}{d_y} - \frac{\triangle + \blacksquare}{d_y}\right)_+ + \frac{\triangle}{d_y} \quad \text{(SM27)}$$

In general, the formula for ORC in the cycle independent mode can be expressed as

$$\begin{aligned}\kappa_{xy} = &-\left(1 - \frac{1}{d_x} - \frac{1}{d_y} - \frac{\triangle}{\min[d_x, d_y]} - \frac{\blacksquare + \blacklozenge}{\max[d_x, d_y]}\right)_+ \\ &-\left(1 - \frac{1}{d_x} - \frac{1}{d_y} - \frac{\triangle + \blacksquare}{\max[d_x, d_y]}\right)_+ + \frac{\triangle}{\max[d_x, d_y]}\end{aligned} \quad \text{(SM28)}$$

We assume that all 4,5-cycles are ineffective, i.e., $\blacksquare = 0 \wedge \blacklozenge = 0$, then Eq. (SM28) can be transformed into Eq. (SM29) that is the lower bound of Eq. (5) [14,20,25].

$$\begin{aligned}\kappa_{xy} = &-\left(1 - \frac{1}{d_x} - \frac{1}{d_y} - \frac{\triangle}{\min[d_x, d_y]}\right)_+ \\ &-\left(1 - \frac{1}{d_x} - \frac{1}{d_y} - \frac{\triangle}{\max[d_x, d_y]}\right)_+ + \frac{\triangle}{\max[d_x, d_y]}\end{aligned} \quad \text{(SM29)}$$

## SM.2 Correlation between the cycle independent mode and CCOM

We assume $d_x \geq d_y$ for convenience. Then, CCOM is formally expressed as Eq. (41).

Based on the definitions in Section 4.2, in the cycle independent mode,

$$F_4 = \frac{\blacksquare}{d_x} \quad \text{(SM30)}$$

$$t = \left(\frac{1}{d_y} - \frac{1}{d_x}\right) \cdot \triangle \quad \text{(SM31)}$$

We analyze $F_5$ in the cycle independent mode as follows:

**Case 1.** $B' \geq 0$

CCOM constructs $\pi'(u, v) = min[m_x(u), m_y(v)] = 1/d_x$ for each 5-cycle $x - u - w - v - y$. Thus, $F_5 = \blacklozenge/d_x$, and Eq. (41) is transformed into

$$\kappa_{xy} \approx -\left(1 - \frac{1}{d_x} - \frac{1}{d_y} - \frac{\triangle}{d_y} - \frac{\blacksquare + \blacklozenge}{d_x}\right)_+ - \left(1 - \frac{1}{d_x} - \frac{1}{d_y} - \frac{\triangle + \blacksquare}{d_x}\right)_+ + \frac{\triangle}{d_x} \quad \text{(SM32)}$$

As is well known, Eq. (SM32) is consistent with Eq. (SM26).

**Case 2.** $B' < 0$

In Case 2, we can confirm that

$$B' = 1 - \frac{1}{d_x} - \frac{1}{d_y} - \frac{\triangle + \blacksquare + \blacklozenge}{d_x} < 0 \quad \text{(SM33)}$$

**Case 2.1** $B' < 0 \wedge (-B' < \blacklozenge/d_x)$

In Case 2.1, we can confirm that

$$\frac{\blacklozenge}{d_x} + B' = 1 - \frac{1}{d_x} - \frac{1}{d_y} - \frac{\triangle + \blacksquare}{d_x} > 0 \quad \text{(SM34)}$$

$$F_5 = \frac{\blacklozenge}{d_x} + B' \quad \text{(SM35)}$$

Based on Eq. (SM33) and Eq. (SM34), Eq. (SM26) is transformed into

$$\kappa_{xy} = -1 + \frac{1}{d_x} + \frac{1}{d_y} + \frac{2}{d_x} \cdot \triangle + \frac{1}{d_x} \cdot \blacksquare \tag{SM36}$$

Based on Eqs. (SM30), (SM31), (SM34) and (SM35), Eq. (41) is transformed into

$$\kappa_{xy} \approx -1 + \frac{1}{d_x} + \frac{1}{d_y} + \frac{2}{d_x} \cdot \triangle + \frac{1}{d_x} \cdot \blacksquare \tag{SM37}$$

As is well known, Eq. (SM37) is consistent with Eq. (SM36).

**Case 2.2** $B' < 0 \wedge (-B' \geq \blacklozenge / d_x)$

In Case 2.2, we can confirm that

$$\frac{\blacklozenge}{d_x} + B' = 1 - \frac{1}{d_x} - \frac{1}{d_y} - \frac{\triangle + \blacksquare}{d_x} \leq 0 \tag{SM38}$$

$$F_5 = 0 \tag{SM39}$$

Based on Eq. (SM33) and Eq. (SM38), Eq. (SM26) is transformed into

$$\kappa_{xy} = \frac{1}{d_x} \cdot \triangle \tag{SM40}$$

Based on Eqs. (SM30), (SM31), (SM38) and (SM39), Eq. (41) is transformed into

$$\kappa_{xy} \approx \frac{1}{d_x} \cdot \triangle \tag{SM41}$$

As is well known, Eq. (SM41) is consistent with Eq. (SM40).

In summary, ORC in the cycle independent mode, i.e., Eq. (SM26) is a particular form of that in the cycle overlap mode, i.e., Eq. (41).

## **Appendix SM.A:** If $A' < 0$, node $y$ has no exclusive neighbors.

$$A' = 1 - \frac{1}{d_x} - \frac{1}{d_y} - \frac{\triangle + \blacksquare + \blacklozenge}{d_y} < 0 \tag{SM.A.1}$$

Based on Eq. (SM.A.1) and $d_x \geq d_y$,

$$\triangle + \blacksquare + \blacklozenge > d_y - 1 - \frac{d_y}{d_x} \geq d_y - 2 \tag{SM.A.2}$$

Because

$$1 + \triangle + \blacksquare + \blacklozenge + |\mathcal{N}_{exc}(y)| = d_y \tag{SM.A.3}$$

We can obtain that

$$\triangle + \blacksquare + \blacklozenge \leq d_y - 1 \tag{SM.A.4}$$

Based on Eq. (SM.A.2) and Eq. (SM.A.4),

$$d_y - 2 < \triangle + \blacksquare + \blacklozenge \leq d_y - 1 \tag{SM.A.5}$$

Because $\triangle + \blacksquare + \blacklozenge$ is an integer, we can obtain that

$$\triangle + \blacksquare + \blacklozenge = d_y - 1 \tag{SM.A.6}$$

Based on Eq. (SM.A.3) and Eq. (SM.A.6), we can obtain that $|\mathcal{N}_{exc}(y)| = 0$. Therefore, node $y$ has no exclusive neighbors.